\documentclass[11pt,letterpaper]{article}
\usepackage{enumerate}
\usepackage{amsthm}
\usepackage{amsmath,latexsym,amssymb,enumerate,graphicx}
\usepackage{hyperref}

\usepackage{fullpage}
\usepackage{color}
\usepackage{cancel}
\usepackage{float}
\newcommand{\nc}{\newcommand}

\newcommand{\<}{\langle}
\renewcommand{\>}{\rangle}

\newtheorem{theorem}{Theorem}
\newtheorem{definition}[theorem]{Definition}
\newtheorem{lemma}[theorem]{Lemma}

\newtheorem{proposition}[theorem]{Proposition}

\newenvironment{mylist}[1]{\begin{list}{}{
	\setlength{\leftmargin}{#1}
	\setlength{\rightmargin}{0mm}
	\setlength{\labelsep}{2mm}
	\setlength{\labelwidth}{8mm}
	\setlength{\itemsep}{0mm}}}
	{\end{list}}

\newcommand{\lemref}[1]{Lemma~\ref{lem:#1}}
\nc\eq[1]{(\ref{eq:#1})}

\newcommand{\NP}{\textsc{NP}}

\newcommand{\MIP}{\textsc{MIP}}
\newcommand{\QMA}{\textsc{QMA}}
\newcommand{\IP}{\textsc{IP}}
\newcommand{\EXP}{\textsc{EXP}}
\newcommand{\NEXP}{\textsc{NEXP}}
\newcommand{\PSPACE}{\textsc{PSPACE}}

\renewcommand{\a}{\alpha}
\newcommand{\h}{\mathcal{H}}
\newcommand{\bra}[1]{\< #1 |}
\newcommand{\ket}[1]{| #1 \>}
\newcommand{\eps}{\epsilon}

\newcommand{\Xlin}{\mathcal{X}} 
\newcommand{\Zlin}{\mathcal{Z}}

\DeclareMathOperator{\E}{\mathbf{E}}
\DeclareMathOperator{\Id}{\mathbb{I}}
\DeclareMathOperator{\poly}{poly}
\DeclareMathOperator{\Tr}{Tr}
\DeclareMathOperator{\CON}{CON}

\renewcommand{\Re}{\mathrm{Re}}

\newcommand{\avg}[2]{\left\langle #1 \right\rangle_{#2}}
\newcommand{\bavg}[2]{\big\langle #1 \big\rangle_{#2}}
\newcommand{\Bavg}[2]{\Big\langle #1 \Big\rangle_{#2}}

\newcommand{\C}{\mathbb{C}}
\newcommand{\R}{\mathbb{R}}
\newcommand{\N}{\mathbb{N}}

\newcommand{\sx}{\hat{X}}
\newcommand{\sz}{\hat{Z}}

\renewcommand{\sp}{\hat{P}}
\newcommand{\cx}{\bar{X}}
\newcommand{\cz}{\bar{Z}}

\begin{document}
\title{Constant-Soundness Interactive Proofs for Local Hamiltonians}
\author{Anand Natarajan\thanks{Center for Theoretical Physics,
    MIT, Cambridge, USA. email:\texttt{anandn@mit.edu}. } \qquad Thomas
  Vidick\thanks{Department of Computing and Mathematical Sciences,
    California Institute of Technology, Pasadena, USA. email:
    \texttt{vidick@cms.caltech.edu}.}} 
\date{\today}
\maketitle

\begin{abstract}
  We give a quantum multiprover interactive proof system for the local
  Hamiltonian problem in which there is a constant number of provers,
  questions are classical of length polynomial in the number of
  qubits, and answers are of constant length.
	The main novelty of our protocol is that the gap between
  completeness and soundness is directly proportional to the promise
  gap on the (normalized) ground state energy of the Hamiltonian. This
  result can be interpreted as a concrete step towards a quantum
  PCP theorem giving entangled-prover interactive proof systems for
  QMA-complete problems. 
	
	The key ingredient is a quantum version of the classical linearity
  test of Blum, Luby, and Rubinfeld, where the function
  $f:\{0,1\}^n\to\{0,1\}$ is replaced by a pair of functions $\Xlin, \Zlin:\{0,1\}^n\to \text{Obs}_d(\C)$, the set of
  $d$-dimensional Hermitian matrices that square to identity. The test enforces that (i) each function is exactly linear, $\Xlin(a)\Xlin(b)=\Xlin(a+b)$ and $\Zlin(a) \Zlin(b)=\Zlin(a+b)$, and (ii) the two functions are approximately complementary, $\Xlin(a)\Zlin(b)\approx (-1)^{a\cdot b} \Zlin(b)\Xlin(a)$. 
\end{abstract}

\section{Introduction}

The theory of NP-completeness is a central part of complexity theory,
and an important area of research in quantum complexity theory over the
last two decades has been to characterize the quantum analog of
NP-completeness. The foundations for this were laid by Kitaev, who
established a quantum version of the Cook-Levin theorem~\cite{KitSheVya02}. His result shows that
the \emph{local Hamiltonian problem} is complete for the complexity
class QMA, the quantum analog of NP. The local Hamiltonian problem can
be cast as a quantum analog of Boolean constraint satisfaction problems (CSPs):
instead of the satisfiability of a formula consisting of a conjunction of
clauses each acting on  a
few Boolean variables, one considers the problem of finding the
minimum eigenvalue of a Hermitian operator (the Hamiltonian)
consisting of the sum of
\emph{local terms}, each acting on a constant number of quantum bits (qubits). Boolean CSPs
are a special case of the local Hamiltonian problem, obtained by restricting all terms in the Hamiltonian to be  matrices diagonal in the computational
basis. The Hamiltonian operator plays a fundamental role in quantum mechanics, and the constraint of locality is motivated by problems considered in many-body physics where physical interactions typically only involve small groups of neighboring particles. Eigenvalues of the Hamiltonian are called energy levels, and the study of the minimal energy (smallest eigenvalue) and associated eigenvector (the ground state) is the focus of condensed-matter physics, as they correspond  to the energy and equilibrium state of the system at zero temperature respectively. The local
Hamiltonian problem therefore provides a compelling abstraction within which to frame many of the computational
problems that arise in fundamental areas of physics.

The NP-hardness of deciding exact satisfiability of Boolean CSPs stated in the Cook-Levin theorem has been greatly strengthened by the PCP theorem~\cite{AroLunMotSudSze98JACM,AroSaf98JACM}, which extends NP-hardness to the problem of \emph{approximating} the maximum satisfiable
fraction of clauses in a formula, even up to \emph{constant factors}. This result forms a cornerstone of modern complexity theory, and in particular implies many optimal (under P$\neq$NP) hardness-of-approximation results for
 NP-complete problems of independent interest~\cite{FeiGolLovSafSze96JACM}. 
The existence of a quantum analog of the PCP theorem, stating QMA-hardness of constant-factor approximations to the minimal energy of a local Hamiltonian $H$ (normalized so that $\|H\|=O(1)$), is a major open problem in quantum complexity theory called the quantum PCP conjecture~\cite{AharonovN02qma,AharonovAV13qpcp}. For reasons that will soon be clear we refer to this conjecture as the ``constraint satisfaction'' variant of QPCP.

The PCP theorem has several different formulations and
proofs; arguably the simplest is a combinatorial proof developed by
Dinur~\cite{Dinur07pcp}. This proof seems challenging to
quantize~\cite{AharonovILV09detectability}; for instance, it
relies heavily on copying bits, but copying qubits is forbidden by the
no-cloning theorem (see~\cite[Section 3]{AharonovAV13qpcp} for a discussion of the many more difficulties that arise). The original proof of the theorem was quite different,
drawing very strongly on the new connections that were being established between the theory of error-correcting codes, testing, and the power of interactive proof
systems~\cite{arora1994probabilistic}. 
A major milestone along this route is the inclusion $\text{NEXP} \subseteq
\text{MIP}$~\cite{BabForLun91CC}, a result that can be interpreted as
a ``scaled up'' precursor to the PCP theorem. A quantum analog of this
result was first suggested as an alternative formulation of (and as a
step towards a proof of) the quantum PCP conjecture in~\cite{FV14}:
does the inclusion $\mathrm{QMA_{EXP}} \subseteq \text{QMIP}^*$
hold?\footnote{Here $\text{QMIP}^*$ denotes the class of languages
  that have multi-prover interactive proofs with a quantum
  polynomial-time verifier and quantum entangled provers;
  $\mathrm{QMA_{EXP}}$ is to $\QMA$ what NEXP is to NP.} We henceforth
refer to this inclusion as the ``multiplayer games'' variant of
QPCP. Although both formulations, ``multiplayer games'' and
``constraint satisfaction'', of the classical PCP theorem are easily
seen to be equivalent, whether a similar equivalence holds in the
quantum world is an interesting open problem; we refer
to~\cite[Section 5.4]{AharonovAV13qpcp} for a more in-depth
discussion.

\subsection{Main result}
We make progress on the ``multiplayer games'' variant of QPCP by considering a quantum analog
of what is often presented as the first step of the proof
of the classical PCP theorem: the (games variant of the) exponentially
long PCP for NP, based on the linearity test of Blum, Luby and
Rubinfeld~\cite{BLR93} (see e.g. Theorem~18.21 in the
book~\cite{AroBar09} for a precise formulation). Specifically, we
consider the question
of proving the inclusion $\QMA \subseteq  \text{MIP}^*$, where the
interactive protocol is restricted to a single round with constant answer length
(but polynomial question length). The analogous statement with $\text{NP}$ and
$\text{MIP}$ in place of $\text{QMA}$ and $\text{MIP}^*$ is exactly
the classical exponentially long PCP.

In fact the inclusion we are seeking \emph{does} hold, and follows from known results in complexity theory. For
instance, combining the trivial inclusions $\QMA\subseteq\EXP\subseteq \NEXP$ with
  $\NEXP\subseteq \MIP^*$~\cite{IV12} together with the fact that the
  latter holds for a protocol involving a single round of interaction
  and a constant answer length~\cite{Vidick13xor} suffices to establish the result. A different, perhaps simpler route would be to use $\QMA\subseteq\PSPACE$ and $\IP=\PSPACE$~\cite{Sha92}, applying
  arithmetization to obtain an $\IP$ protocol and then introducing a second
  prover and the technique of~\cite{KKMV09} to parallelize the protocol to a single
  round of interaction with two entangled provers; finally one would have to apply some form of parallel amplification~\cite{KV11parallel,bavarian2015anchoring} technique and finish with a method that allows a reduction in the answer length~\cite{Vidick13xor}. 
	
	Working through the reductions implied by either of these routes leads to a complex
  protocol in which the structure of the original instance of the
  local Hamiltonian problem has all but disappeared, and the ``proof''
  held by the provers bears little relation to the ground
  state of the local Hamiltonian --- even though it still, of course,
  suffices to certify its ground state energy. 

In this work we provide a simpler, more direct and arguably ``more quantum'' construction of an ``exponentially-long quantum PCP''.  The key ingredient of our protocol is a 
quantum generalization of the BLR linearity test --- a ``truly quantum'' generalization in the sense that honest provers are \emph{required} to apply quantum operations on a shared entangled state in order to achive completeness. (This is in contrast to the entangled-prover linearity test of~\cite{IV12}, where the entanglement between the provers is  treated as a hurdle against which the protocol has to be ``immunized''.)
 Moreover, our protocol uses the
structure of the local Hamiltonian problem in a natural way: honest provers are asked
  to share a distributed encoding of the ground state, on which they
  perform measurements in order to determine their
  answers. Theorem~\ref{thm:main0} gives a precise statement of our result.

\begin{theorem}\label{thm:main0}
Let $H$ be a local Hamiltonian on $n$ qubits such that $\|H\|\leq 1$, and 
$\lambda_{min}\in [0,1]$  its smallest eigenvalue. Let 
 $0 \leq a(n) < b(n) \leq 1$ be such that $b(n) - a(n) \geq 1 /
\poly(n)$. Then there
exists an interactive proof system between a classical polynomial-time
verifier and seven entangled provers that decides
whether $\lambda_{min} \leq a(n)$ or $\lambda_{min} \geq b(n)$ with completeness $2/3$
and soundness $1/3$. The proof system involves a single
round of interaction in which the verifier sends $\poly(n)$
bits to each prover and receives $O(1)$ bits from each.
\end{theorem}

	Beyond the statement of the theorem itself,  we believe our proof technique is an important
  contribution in the quest for a quantum PCP theorem, either in
  its ``constraint satisfaction'' or ``multiplayer games'' variants. The conjecture is a distant target, and its resolution will undoubtedly require
many new ideas. Although the two proofs of the
classical PCP theorem provide valuable starting points, it also seems important to find
the ``right'' quantum generalizations of the main ingredients ---
possibly requiring altogether new ones in order to overcome the
specific obstacles posed by quantum information (see
e.g.~\cite[Section 3]{AharonovAV13qpcp}).	Our results mark a step in this programme. To describe them further we introduce our quantum linearity
test next, and then explain its use in devising a protocol for the
proof of Theorem~\ref{thm:main0}.
	
\subsection{A quantum linearity test}

What is a good quantum analogue of the linearity test? Recall that in our context instead of evaluating \emph{clauses} on \emph{Boolean variables} we are interested in evaluating \emph{local Hamiltonians} on \emph{qubits}. In addition we may without loss of generality assume that each local term is a tensor product of single-qubit Pauli operators $\mathcal{P}=\{I,X,Z\}$ (defined in~\eqref{eq:pauli-def}), as the corresponding local Hamiltonian problem is known to be QMA-complete~\cite{CM13}. 

Thus a natural starting point consists in replacing the
Boolean function $f:\{0,1\}^n\to\{0,1\}$ by a function $\hat{F}:\{0,1\}^n \to \mathcal{P}^{\otimes n}$. Linearity would then amount to the requirement that $\hat{F}(a + b)=\hat{F}(a)\hat{F}(b)$ for every $a,b\in \{0,1\}^n$, where the underlying operation is matrix multiplication. Can such a property be tested? 

Our quantum linearity test provides a positive answer for a specific instantiation of the question that is the most appropriate to our setting. The test considers \emph{pairs} of functions
 $\sx,\sz:\{0,1\}^n \to \mathcal{O}$, where $\mathcal{O}$ denotes the set of all observables (Hermitian matrices that square to identity), \emph{of any dimension}: indeed, we do not impose a priori that the functions take values in the set $\mathcal{P}^{\otimes n}$ of $n$-qubit Pauli operators. The goal of the test is to enforce, to the extent possible, that $\sx(a) \simeq X^{a_1}\otimes\cdots \otimes X^{a_n}$ and $\sz(b) \simeq Z^{b_1}\otimes\cdots \otimes Z^{b_n}$ for all $a,b\in\{0,1\}^n$, where $\simeq$ denotes ``behaves as'', in a sense soon to be made precise. 

We design a test for this property. The test can be implemented by a verifier interacting classically with $r$ entangled provers, where $r$ is a parameter of the test related to the use of an $r$-qubit quantum stabilizer code; we can take $r=7$. Each of the provers is sent a query of the form $(X,a)$ or $(Z,b)$, where $X,Z$ are treated as formal labels and $a,b\in \{0,1\}^n$, and replies with a single $\pm 1$ bit.\footnote{In fact the test considers an two additional types of queries, with two-bit answers; see Section~\ref{sec:protocol} for details.} We show that the test has the following completeness and soundness properties (see Lemma~\ref{lem:quantum-linearity} for a formal statement):
\begin{mylist}{\parindent}
\item[(\emph{Completeness})] If the provers each own one of the $r$ shares of each of the $n$ qubits of an arbitrary $n$-qubit state $\ket{\psi}$, encoded via an $r$-qubit stabilizer code specified in the protocol, and measure their respective share using observable $X(a) = X^{a_1}\otimes\cdots\otimes X^{a_n}$ (resp. $Z(b) = Z^{b_1}\otimes\cdots \otimes Z^{b_n}$) when queried with $(X,a)$ (resp. $(Z,b)$) then they succeed in the test with probability $\omega^*_{encode}$ (a universal constant). 
\item[(\emph{Soundness})] For any $\eps>0$ and strategy of the provers which succeeds with
  probability at least $\omega^*_{encode}-\eps$ in the test, for each prover there exists
  $\Xlin,\Zlin:\{0,1\}^n\to\mathcal{O}$, where $\mathcal{O}$ is the set of observables acting on the prover's Hilbert space, such that
	\begin{mylist}{2em}
	\item[(i)] $\Xlin$ and $\Zlin$ are \emph{exactly linear}, in the sense that $\Xlin(a+b)=\Xlin(a)\Xlin(b)$ for all $a,b\in\{0,1\}^n$ and similarly for $\Zlin$;
	\item[(ii)] The action of $\Xlin$ and $\Zlin$ on the provers' shared entangled state is indistinguishable, up to an additive $O(\eps^{1/16})$ in Euclidean norm, from the action of the provers' actual observables $\sx, \sz$ in their  strategy; 
	\item[(iii)] There exists a local isometry acting on each provers' local Hilbert space such that the action of $\Xlin(a)$, $\Zlin(b)$ 
 on the isometry's input (the provers' shared entangled state in the strategy) is (up to $O(\eps^{1/16})$ in Euclidean norm) equivalent to the action of the Pauli operators $X(a),Z(b)$ on the isometry's
  ouptut. 
  %Moreover, the isometry always produces an $(rn)$-qubit encoding of an $n$-qubit entangled state, where each group of $r$ qubits lies in the codespace of the stabilizer code employed.
	\end{mylist}
\end{mylist}

A key feature of the test, essential to obtaining a gap-preserving reduction, is that $n$ does not appear in the soundness bound --- in the language of testing, the test has \emph{constant completeness-soundness gap}. (In addition the test enforces a number of useful consistency properties between the provers' operations that are used to prove our main theorem; we refer to Section~\ref{sec:analysis} for details.)
The actual test (see the three \emph{encoding tests} in Figure~\ref{fig:protocol} for a description) and its analysis combine two main ingredients. 

The first ingredient is the entangled-prover
linearity test of~\cite{IV12}, adapted into a two-prover test using the oracularization technique from~\cite{IKM09}. This test is used to verify that the provers' observables associated with $X$ or $Z$-type queries, represented by functions $\sx$ and $\sz$, are close to linear \emph{when considered separately}. The analysis of the test follows the arguments from~\cite{IV12,IKM09}. We make the important observation that 
property (i) of \emph{exact linearity} described above can be guaranteed to hold at the level of the observables themselves for a pair of functions $\Xlin$, $\Zlin$ that are close to the provers' $\sx$ and $\sz$. The property, although already implicit in~\cite{IV12,IKM09}, did not play a significant role in their analysis. For us it is crucial, as it allows certain relations to hold with \emph{zero error}, at the level of operators rather than being state-dependent (for an example where the property is needed, see the beginning of the proof of Lemma~\ref{lem:iso-single-prover-xz}). 

For our purposes the linearity obtained from the entangled-prover linearity test alone is far from sufficient, and indeed there is nothing ``quantum'' about it: provers not sharing any entanglement, and replying deterministically, can of course succeed in the test as long as their answers are given according to a pair of linear functions. 

Therefore a second ingredient is used to turn the classical linearity test above into a ``quantum'' linearity. This makes use of the uniquely quantum ``complementarity'' between $X$ and $Z$ Pauli observables: the two operators anti-commute. This is exploited by the famous \emph{CHSH test}, a test that can only be successfully passed by provers sharing entanglement. Here we follow closely the work of~\cite{ji2015classical} and combine the CHSH test with a ``stabilizer test'' that relies on certain properties of the code used to distribute qubits between the provers (of course malicious provers need not a priori be using this encoding at all --- it is a consequence of the test that they should). This part of the test could not be passed by provers who do not share any entanglement: there is no
 classical randomized strategy that enables the provers to simultaneously sample 
from the output distribution generated by the application of $\sx(a)$ and $\sz(b)$ on a code state, as required by the protocol.\footnote{It is an interesting question
whether the test can be used to verify a high degree of entanglement
between the provers; see Section~\ref{sec:future-work} for further discussion.}

\subsection{An exponential quantum PCP}

We explain the role played by the quantum linearity test in the proof of Theorem~\ref{thm:main0}.
Our starting point are the results~\cite{FV14,ji2015classical}, which provide a ``multiplayer games'' analog of the Cook-Levin theorem for QMA-complete problems (in contrast to Kitaev's theorem, which provides the ``constraint satisfaction'' analog). In particular, Ji~\cite{ji2015classical} gives a five-prover one-round classical interactive proof system for the local Hamiltonian problem such that the verifier's maximum acceptance
probability is $1-K\lambda_{min}(H)n^{-\kappa}$ for constants $K,\kappa$ (see~\cite[Theorem 23]{ji2015classical} for a precise statement). This is sufficient to establish an inverse-polynomial soundness-completeness gap for instances whose minimum eigenvalue are separated by an inverse polynomial. 

Moving forward, the most direct route to a proof of the ``multiplayer games'' variant of the quantum PCP conjecture, i.e. an interactive proof system for the local Hamiltonian problem with \emph{constant} completeness-soundness gap, faces at least two substantial difficulties. 
First, one should establish a \emph{gap-preserving} reduction, whereby the verifier's maximum acceptance probability is related to $\lambda_{min}(H)$ up to constant, instead of inverse polynomial, factors. 
Second, it appears like one would still need to establish a ``constraint satisfaction'' variant of the quantum PCP conjecture in order to allow the gap-preserving reduction to start from a family of instances of the local Hamiltonian problem for which constant approximations to the minimum energy are QMA-hard. Strictly speaking however this step may not be needed, as the transformation provided in the first step may ultimately be \emph{gap-introducing}, as is the case in e.g. Dinur's proof of the classical PCP theorem.   

Our quantum linearity test provides a step towards resolving the first obstacle: we use it to derive a gap-preserving reduction, albeit at the cost of an exponential blow-up in question length. Indeed, Theorem~\ref{thm:main0} relies on a protocol with questions of length polynomial in the number of qubits, or variables, of the local Hamiltonian instance, where ideally the \emph{number} of questions would be polynomial (and their length logarithmic).\footnote{It is this exponential blow-up which makes Theorem~\ref{thm:main0} follow from results in the literature on classical and quantum interactive proofs, as already mentioned. See Section~\ref{sec:future-work} below for further discussion.}
 Taking stock of this loss, however, allows us to provide a simple solution to the second obstacle, thereby providing an unconditional constant-gap interactive proof system for the local Hamiltonian problem. Our method is straightforward: we expand an initial inverse-polynomial promise gap on the smallest eigenvalue by taking appropriate tensor powers of the Hamiltonian.\footnote{We emphasize that the main contribution of our work, and the respect in which it provides a \emph{gap-preserving} reduction, is \emph{not} due to this simple \emph{gap-amplification} trick. Rather, it is in our quantum linearity test and accompanying interactive proof system which provide a gap-preserving reduction irrespective of whether amplification has been performed or not.} While this destroys the locality of the Hamiltonian, it is not an issue for us as our protocol is able to handle any Hamiltonian that is a linear combination of terms made of the tensor product of an arbitrary number of $X$ and $Z$ Pauli operators.\footnote{In order for the verifier to remain polynomial-time we do need to be able to sample from the distribution implied by the modulus of the corresponding coefficients, which in general precludes starting from an arbitrary Hamiltonian and expanding it in the Pauli basis. This is not a problem for the amplification procedure described here.} 

In order to connect the provers' answers, when asked to perform measurements involving both Pauli $X$ and $Z$ operators, to the conclusion of the quantum linearity test, which characterizes their operations on queries that involve only $X$ or $Z$ operators separately, we introduce an additional \emph{consistency test}. In this test one of the provers is asked to measure a term involving both type of Paulis, while the others are asked for a single type. The results are then checked for consistency, providing the desired consistency between different types of queries. 

An interesting feature of our protocol, already implicit in~\cite{ji2015classical}, arises from the use of the stabilizer encoding that underlies the protocol. As already mentioned the $r$ honest provers should distribute an arbitrary state between themselves by encoding it one qubit at a time using an $r$-qubit stabilizer code. This allows them to operate jointly on the encoded state by applying any \emph{transversal gate}, which are logical gates that can be implemented by performing a local operation on each of the encoded qubits. Pauli $X$ and $Z$ unitaries are transversal gates for the code we employ, but more complicated codes can support other types of gates. Thus using entanglement between the provers the verifier is able to ``orchestrate'' certain operations on the entangled state without any prover knowing what gate is being performed (as two different logical gates may in general be implemented transversally with a subset of the provers still performing the same operation in both cases). In our analysis the property is leveraged by treating the $r$ ``physical'' provers as two ``logical'' provers, each consisting of a subset of the physical provers. This effectively lets us formulate, and analyze, most of the protocol as being performed with two provers only; see Section~\ref{sec:protocol} for more details.

\subsection{Related work}
\label{sec:related}

We build on a number of previous works in quantum information
and complexity theory. 

First we mention that motivation for the problem we consider goes back to a question of Aharonov and Ben-Or (personal communication, 2013), who asked how a quantum generalization of the exponential classical PCP could look like if it was not derived through the ``circuitous route'' obtained as the compilation of known but complex results from the theory of classical and quantum interactive proof systems (as described earlier). In this respect we point to~\cite[Section
  5]{AharonovAV13qpcp} for a very different approach to the same question based on a ``quantum take'' on the arithmetization technique. 
	
More directly, our work builds on the already-mentioned 
works~\cite{FV14, ji2015classical} initiating the study of 
entangled-prover interactive proof systems for the local Hamiltonian problem. The idea of using a distributed encoding of the ground
  state in order to obtain a multiprover interactive proof system for
  the ground state energy is introduced in~\cite{FV14}. In that work the protocol required the provers to return qubits; the possibility for making the protocol purely classical was uncovered by Ji~\cite{ji2015classical}. Our use of stabilizer codes, and the stabilizer test which forms part of our protocol, originate in his work. In addition we build upon ideas introduced in the study of 
quantum multiprover interactive proofs with entangled provers~\cite{KM03, CHTW04}, and especially the three-prover linearity test of~\cite{IV12} and the use of oracularization from~\cite{IKM09} to make it into a two-prover test. Finally we draw upon important results from the quantum self-testing literature; in
particular, self tests for the graph states~\cite{McKague13}. 
		
Compared to the works mentioned above, and~\cite{FV14, ji2015classical} in particular, our result differs in two important
  respects, making it incomparable in general. First, the question
  size in our protocol is much larger: $\poly(n)$ bits instead of $O(\log
  n)$ for~\cite{ji2015classical}. Second, the dependence of the
  verifier's acceptance probability on the ground state energy is much
  better: while our dependence is of a constant factor,
  in~\cite{ji2015classical} there is a polynomial
  scaling.\footnote{One could attempt to recover our result by
    repeating the protocol in~\cite{ji2015classical} a polynomial
    number of times. Provided there existed an appropriate parallel
    repetition theorem, this would amplify the soundness to a
    constant. However, the answer length would now be polynomial, and
    it is unclear whether this could be reduced to a constant without
    having to go once more through the complicated reductions
    of~\cite{Vidick13xor}, defeating the purpose.}  
Interpreting all three results as steps towards a quantum PCP
theorem,~\cite{FV14,ji2015classical} propose a first step that is
\emph{size-preserving} (the number of questions is polynomial in the
instance size) but has only an inverse polynomial gap; in contrast we
take the route of a \emph{gap-preserving} construction, but the number
of questions becomes exponential in the instance size. 

Our results are also related to work in quantum property testing~\cite{montanaro2013survey}, and in particular testing EPR pairs~\cite{McKagueYS12rigidity} and more general entangled states~\cite{mckague2014self}. In this setting state-of-the-art results~\cite{ReichardtUV13nature} essentially show how the presence of $n$ EPR pairs between two provers can be certified via a protocol using queries and answers of length polynomial in $n$, with inverse-polynomial completeness-soundness gap. Thus here again no ``constant-gap'' results are known, where the gap would remain constant as the number of  EPR pairs tested grows. Our work is incomparable: our protocol has constant gap but \emph{does not} by itself suffice to certify that the provers share a large number of EPR pairs that are in \emph{tensor product} form. 

Very recently and independently of our work, McKague~\cite{McKague15} has proposed protocols for self-testing many-qubit states that achieve a guarantee similar to ours, i.e. the protocol certifies that there exists an isometry acting on the provers' state for which the expectation values of Pauli operators on the output are close to the expectation values of the provers' measurements (see e.g. Lemma~\ref{lem:iso-single-prover}). However, his protocol is not directly comparable to ours since it requires $\poly(n)$-bit answers and $\log(n)$-bit questions, whereas we use $O(1)$-bit answers and $\poly(n)$-bit questions; in addition and more importantly for us the completeness-soundness gap in his protocol scales polynomially with $n$, whereas in our case the scaling is independent of $n$.

%  Games PCP: Fitzsimons Vidick, Ji
%  Delegated computation: McKague, Broadbent, RUV
%  Quantum games hardness: IKM, Ito Vidick, etc
%  Classical: Babai Fortnow Lund, Blum Luby Rubinfeld

% PCP: Harrow \& Eldar. Also uses the Hadamard code as a locally
% testable code to build on. Not clear if this connection is very
% meaningful. 

\subsection{Directions for future work}
\label{sec:future-work}

 Improving the question length from polynomial in $n$ (in fact, linear in $n$: the polynomial dependency only enters our result through the amplification procedure, but is not needed for the gap-preserving reduction itself) to linear
  in $\log n$ would give a proof of the ``multiplayer games'' variant of the quantum PCP conjecture stated in~\cite{FV14}. We expect
  this to present a significant challenge (note that it would recover,
  and strengthen, the inclusion $\NEXP\subseteq\MIP^*$), but it forms the
  motivation behind our work.  In the classical setting, the key
  ingredient in the proof of $\NEXP \subseteq \MIP$ consists in replacing the
  linearity test with a test for
  \emph{multilinear} functions, or more generally for low-degree
  multivariate polynomials. There has also been recent work in the
  context of direct-sum testing, which directly achieves a linearity
  test with reduced query length~\cite{DDGKS14}. It is an interesting open question
  whether these tests can be generalized to the quantum setting, extending our quantum linearity test. 
	
	More
  generally, the area of \emph{device-independent} quantum property
  testing has many interesting open problems~\cite{montanaro2013survey}, which however
  almost systematically suffer from inverse-polynomial
  completeness-soundness gaps as soon as the property tested scales in
  size. Our results may suggest novel approaches to some of these
  problems.

\paragraph{Organization of the paper.}
In Section~\ref{sec:prelim} we introduce some notation used throughout as well as basic definitions on stabilizer codes and local Hamiltonians. In Section~\ref{sec:protocol} we describe the protocol used for the proof of Theorem~\ref{thm:main0}. In Section~\ref{sec:analysis} we analyze the quantum linearity test performed as part of the protocol as a stand-alone test. In Section~\ref{sec:game-lh} we conclude the analysis of the protocol, leading to the proof of Theorem~\ref{thm:main0}. 

%----------------------%
\section{Preliminaries}
\label{sec:prelim}
%----------------------%

We assume basic familiarity with quantum information but give all required definitions. We refer to the standard textbook~\cite{NieChu01} for additional background material.

\subsection{Quantum states and measurements}

A $n$-qubit quantum state is represented by a unit vector $\ket{\psi}\in \C^2 \otimes \cdots \otimes \C^2 = (\C^2)^{\otimes n} \approx \C^{2^n}$, where the ket notation $\ket{\cdot}$ is used to signify a column vector. A bra $\bra{\psi}$ is used for the conjugate-transpose $\bra{\psi} = \ket{\psi}^\dagger$, which is a row vector. We use $\|\ket{\psi}\|^2 = |\bra{\psi}\psi\rangle|$ to denote the Euclidean norm, where $\bra{\psi}\phi\rangle$ is the skew-Hermitian inner product between vectors $\ket{\phi}$ and $\ket{\psi}$. For a matrix $X$, $\|X\|$ will refer to the operator norm, the largest singular value. When the Hilbert space can be decomposed as $\h = \h_A \otimes \h_B$ for some $\h_A$ and $\h_B$, and $X$ is an operator on $\h_A$, we often write $X$ as well for the operator $X\otimes\Id_{\h_B}$ on $\h$. It will always be clear from context which space an operator acts on. 

A density matrix on $n$ qubits is a positive semi-definite matrix $\rho \in \C^{2^n} \times \C^{2^n}$ of trace $1$. The density matrix associated to $\ket{\psi}$ is the rank-1 projection $\ket{\psi}\bra{\psi}$. 

A $n$-qubit measurement (also called POVM, for projective operator-valued measurement) with $k$ outcomes is specified by $k$ positive matrices $M=\{M_1,\ldots,M_k\}$ in $\C^{2^n}\times\C^{2^n}$ such that $\sum_i M_i = \Id$. The measurement is \emph{projective} if each $M_i$ is a projector, i.e. $M_i^2 = M_i$. The probability of obtaining the $i$-th outcome when measuring state $\rho$ with $M$ is $\Tr(M_i \rho)$. By Naimark's dilation theorem, any POVM can be simulated by a projective measurement acting on an enlarged state; that is, for every POVM $M = \{M_i\}_i$ acting on state $\ket{\psi} \in \mathcal{H}$ there exists a projective measurement $M' = \{P_i\}_i$ and a state $\ket{\psi}\otimes \ket{\phi} \in \mathcal{H} \otimes \mathcal{H}_{\text{ancilla}}$ with the same outcome probabilities as $M$. Moreover, the post-measurement state after performing $M$ is the same as the \emph{reduced} post-measurement state obtained after performing $M'$ and tracing out the ancilla subsystem $\mathcal{H}_{\text{ancilla}}$.

An $n$-qubit observable is a Hermitian matrix $O\in \C^{2^n}\times\C^{2^n}$ that squares to identity. $O$ is diagonalizable with eigenvalues $\pm 1$, $O=P_+-P_-$, and $P=\{P_+,P_-\}$ is a projective measurement. For any state $\rho$, $\Tr(O\rho)$ is the expectation of the $\pm 1$ outcome  obtained when measuring $\rho$ with $P$. If $\rho = \ket{\psi}\bra{\psi}$ we abbreviate this quantity, $\Tr(O\rho) = \Tr(P_+\rho)-\Tr(P_-\rho) = \langle
\psi | O | \psi \rangle$ as $\avg{P}{\psi}$.

A convenient orthogonal basis for the real vector space of $n$-qubit observables is given by the set $\{I,X,Y,Z\}^{\otimes n}$, where $\{I,X,Y,Z\}$ are the four single-qubit Pauli observables
\begin{equation}\label{eq:pauli-def}
 I = \begin{pmatrix} 1 & 0 \\ 0 & 1 \end{pmatrix},\quad
X= \begin{pmatrix} 0 & 1 \\ 1 & 0 \end{pmatrix}, \quad Y = \begin{pmatrix} 0
  & -i \\ i & 0\end{pmatrix}, \quad Z = \begin{pmatrix} 1 & 0 \\ 0 &
  -1\end{pmatrix} .
	\end{equation}
We 
often consider operators that are tensor products of just
$I$ and $X$, or just $I$ and $Z$. We denote these by $X(a), Z(b)$, where
the strings $a, b \in \{0, 1\}^n$ indicate which qubits to apply the
$X$ or $Z$ operators to: a $0$ in position $i$ indicates an $I$ on
qubit $i$, and a $1$ indicates an $X$ or $Z$.

\subsection{Stabilizer codes}
\label{sec:stabilizer}

Stabilizer codes are the quantum analogue of linear codes. For an introduction to the theory of stabilizer codes we refer to~\cite{Gottesman97}. We will only use very elementary properties of such codes. 

The codes we consider are \emph{Calderbank-Shor-Steane (CSS)
  codes}~\cite{CalderbankShor96,Steane96}. For an $r$-qubit code the
codespace, the vector space of all valid codewords, is the subspace of
$(\C^2)^{\otimes r}$ that is the simultaneous $+1$ eigenspace of a set
$\{S_1,\ldots,S_k\}$ of $r$-qubit pairwise commuting Pauli observables
called the stabilizers of the code. The stabilizers form a group under
multiplication. Unitary operations, such as a Pauli $X$ or $Z$
operators, on the logical qubit are implemented on the codespace by
logical operators $X_{logical}$ and $Z_{logical}$. The smallest CSS code is Steane's $7$-qubit code~\cite{Steane96}. Table~\ref{tab:stabilizer} lists a set of stabilizers that generate the stabilizer group of the code. 

\begin{table}
  \centering
  \begin{tabular}[h]{l | c  c  c c c c r }
    &1 & 2 & 3 & 4 & 5 & 6 & 7 \\ \hline
    Stabilizers &I & I & I & X & X  & X & X \\
    &I & X & X & I & I & X & X \\
    &X & I & X & I & X & I & X \\
    &I & I & I & Z & Z  & Z & Z \\
    &I & Z & Z & I & I & Z & Z \\
    &Z & I & Z & I & Z & I & Z \\ \hline
    Logical X & X & X & X & X & X & X & X \\
    Logical Z & Z & Z & Z & Z & Z & Z & Z
  \end{tabular}
  \caption{Stabilizer table for the 7-qubit Steane code}
  \label{tab:stabilizer}
\end{table}
		
Every CSS code satisfies certain properties which will be useful for
us. Firstly, both the stabilizer generators and the logical operators
can be written as tensor products of only $I$, $X$, and $Z$ operators --- there are no
$Y$. This simplifies our protocol, allowing us to consider only two
distinct basis settings. Secondly, every CSS code has the following symmetry: for every index $i \in [r]$ there
exists stabilizers $S_X$, $S_Z$ such that $S_X$ is a tensor product of
only $X$ and $I$ operators and has an $X$ at position
$i$, and $S_Z$ is equal to $S_X$ with all $X$ operators replaced by
$Z$ operators. 

These properties imply the following simple observation,
which will be important for us. For every Pauli operator $P\in\{I,X,Z\}$ acting on the $i$-th qubit of the code there is a tensor product $\bar{P}$ of Paulis acting on the remaining $(r - 1)$ qubits such that $P \otimes \bar{P}$ is a stabilizer
operator on the whole state, and moreover each term in the tensor
product is either identity or $P$. Indeed, the choice of $\bar{P}$ is not unique. Henceforth, we use the
notion $\bar{P}$ to denote \emph{any} such operator, unless
otherwise specified. 

\subsection{Local Hamiltonians}\label{sec:local-h}

A $n$-qubit local Hamiltonian is a Hermitian, positive semidefinite  operator $H$ on $(\C^2)^{\otimes {n}}$ that can be decomposed as a sum $H = \sum_{i=1}^m H_i$ with each $H_i$ is local, i.e. $H_i$ can be written as $H_i = I\otimes\cdots I \otimes h_i \otimes I \otimes \cdots \otimes I$, where $h_i$ is a Hermitian operator on $(\C^2)^{\otimes k}$ with norm (largest singular value) at most $1$. The smallest $k$ for which $H$ admits such a decomposition is called the locality of $H$. The terms are normalized such that $\| H_i\| \leq 1$ for all $i$. A family of Hamiltonians $\{H_i\}$ acting on increasing numbers of qubits is called local if all $H_i$ are $k$-local for some $k$ independent of $n$ (for us $k$ will always be $2$).

The local Hamiltonian problem is the prototypical $\QMA$-complete problem, as 3SAT is for $\NP$. 

\begin{definition}
Let $k\geq 2$ be an integer. 
  The $k$-\emph{local Hamiltonian problem} is to decide, given a
  family of $k$-local Hamiltonians $\{H_n\}_{n\in\N}$ such that $H_n$ acts on $n$ qubits, and functions $a,b:\N\to (0,1)$ such that $b - a = \Omega( \poly^{-1}(n))$, if the smallest eigenvalue
  of $H_n$ is less than $a(n)$ or greater than $b(n)$.
\end{definition}

Here we restrict our attention to Hamiltonians 
\[ H = \frac{1}{m} \sum_{i=1}^{m} H_i, \]
for which each term $H_i$ can be written as a linear combination of tensor products of Pauli $X$
and $Z$ observables only (no $Y$). Such Hamiltonians are known to be QMA
complete; in particular we consider a restricted class of $2$-local Hamiltonians, called Hamiltonians of $XZ$ form, for which each $H_i$ can be written as $H_i = \alpha_{i_1 i_2}(X_{i_1} \otimes X_{i_2} + Z_{i_1} \otimes Z_{i_2})$, where $i_1,i_2\in\{1,\ldots,n\}$ indicate the qubits on which a Pauli $X$ or $Z$ acts and the coefficients $\alpha_{i_1i_2}\in\R$ satisfy $|\alpha_{i_1i_2}|\leq 1$.

\begin{theorem}[Cubitt and Montanaro~\cite{CM13}, lemma
  21]
The local Hamiltonian problem for $XZ$-Hamiltonians is QMA-complete.\footnote{In~\cite[Lemma 21]{CM13} this is stated for the $XY$ Hamiltonian, to which $XZ$ is equivalent by local rotation.}
\end{theorem}

\subsection{State-dependent distance measure}

We make extensive use of a state-dependent distance
between measurements that has been frequently used in the context of entangled-prover interactive proof systems (see e.g.~\cite{IV12,ji2015classical}). Let $\{M^a\}$ and $\{N^a\}$ be two POVMs
with the same number of possible outcomes, indexed by $a$, and let
$\ket{\psi}$ be a quantum state. The \emph{state-dependent distance} between
$M$ and $N$ on $\ket{\psi}$ is defined as
\begin{align*}
  d_\psi(M, N) = \Big(\sum_a \big\|\sqrt{M^a} \ket{\psi} - \sqrt{N^a}\ket{\psi}\big\|^2\Big)^{1/2}.
\end{align*}
To simplify the notation, let $A^a = \sqrt{M^a}$ and $B^a =
\sqrt{N^a}$. Then this distance can be rewritten as:
\begin{align*}
  d_\psi(M, N)^2 &= \sum_a \| A^a \ket{\psi} - B^a  \ket{\psi}\|^2\\
               &=\sum_a \bra{\psi} (A^a - B^a)^2 \ket{\psi} \\
               &= \sum_a \big( \| A^a \ket{\psi} \|^2 + \| B^a \ket{\psi} \|^2 -
                 \bra{\psi} (A^a B^a + B^a A^a) \ket{\psi} \big) \\
               &= 2 - \sum_a \bra{\psi}(A^a B^a + B^a A^a) \ket{\psi}
  \\
               &= 2 - 2\sum_a \Re\big(\bra{\psi} A^a B^a \ket{\psi}\big).
\end{align*}
In the last line we have used the fact that $A^a$ and $B^a$ are
Hermitian. If we specialize to the case of projective measurements
with binary outcomes, we get the following relations (here $A =
A^0 - A^1$ and $B = B^0 - B^1$
are the observables associated to the measurements):
\begin{align}
d_\psi(M, N)^2  &= 2 - \bra{\psi} (A^0 B^0 + A^1 B^1 + B^0 A^0 + B^1
                 A^1) \ket{\psi} \nonumber \\
               &= 2 - \frac{1}{4} \bra{\psi} ((\Id + A)(\Id + B)
                 + (\Id - A)(\Id - B) + (\Id + B)(\Id + A)
                 + (\Id  - B)(\Id - A)) \ket{\psi} \nonumber \\
               &= 2 - \frac{1}{4} \bra{\psi} (4\Id + 2 AB + 2 BA)
                 \ket{\psi} \nonumber \\
               &= 1 - \frac{1}{2} \bra{\psi} (AB + BA) \ket{\psi} \nonumber \\
               &= \frac{1}{2} \bra{\psi} (A - B)^2 \ket{\psi}.\label{eq:dist_observables}
\end{align}
This distance measure has the following useful property:

\begin{lemma}
  Let $\ket{\psi}$ be a quantum state, $\{C_a\}$ a family of operators such that $\|\sum_a C_aC_a^\dagger\|\leq K$ and $\{M^a\}$
  and $\{N^a\}$ POVMs. Then 
  $$\Big|\sum_a \bra{\psi} C_a \sqrt{M^a}\ket{\psi} - \sum_a \bra{\psi} C_a \sqrt{N^a}
    \ket{\psi}\Big| \leq \sqrt{K}d_\psi(M, N).$$
  \label{lem:approx}
\end{lemma}

\begin{proof}
Let $A^a = \sqrt{M^a}$ and $B^a = \sqrt{N^a}$. Applying the Cauchy-Schwarz inequality,
  \begin{align*}
    \Big|\sum_a \bra{\psi} C_a (A^a - B^a) \ket{\psi} \Big| 
    &\leq     \Big|\bra{\psi}\sum_a  C_aC_a^\dagger  \ket{\psi} \Big|^{1/2} \Big|\bra{\psi}\sum_a (A^a -      B^a)^2     \ket{\psi} \Big|^{1/2} \\
    &\leq \sqrt{K} d_\psi(M, N),
  \end{align*}
	as claimed.
\end{proof}
A second measure of proximity that is often convenient is the
\emph{consistency}. As before, let $\{M^a\}$ and $\{N^a\}$ be POVMs
with the same number of outcomes. Then their consistency is defined
as
\[ \CON_\psi(M, N) = \Re \Big(\sum_a \bra{\psi} M^a N^a \ket{\psi}\Big). \]

The following lemma relates the consistency and the state-dependent
distance.
\begin{lemma}[Lemma~10 in~\cite{ji2015classical}]
  Let $\psi$ be a state and $\{M^a\}$ and $\{N^a\}$ be two POVMs with equal numbers of
  outcomes. If $\CON_\psi(M, N) = 1 - \delta$ for some $\delta\geq 0$ then $d_\psi(M,
  N) \leq \sqrt{\delta}$.
  \label{lem:condist}
\end{lemma}

A useful property of the consistency is that if $M$ and $N$ are POVMs acting on two separate subsystems of $\ket{\psi}$, applying Naimark dilation to each of them results in projective measurements $M'$ and $N'$ and a state $\ket{\psi'}$ such that $\CON_\psi(M, N) = \CON_{\psi'}(M', N')$. 

%----------------------%
\section{Description of the protocol}
\label{sec:protocol}
%----------------------%

In this section we describe the protocol used in the proof of Theorem~\ref{thm:main0}. The input to the protocol is an $n$-qubit local Hamiltonian $H$ in XZ form, as described in Section~\ref{sec:local-h}. The verifier interacts with $r$ provers. One should think of ``honest'' provers as sharing a qubit-by-qubit encoding of the  $n$-qubit ground state $\ket{\Gamma}$ of $H$ according to a CSS code, such as Steane's $7$-qubit code in which case $r=7$, as described in Section~\ref{sec:stabilizer}, and of performing the Pauli measurement indicated by the query (a complete description of the honest strategy is given in Definition~\ref{def:honest-strategy}). 

In the protocol, the verifier asks the provers to perform one of a
series of tests chosen according to some pre-specified distribution. In each test the
prover must answer with one or two answer bits, which are encoded as
$\pm 1$ to match the convention that quantum observables
have $\pm 1$ eigenvalues. There are several
possible types of queries that each prover may receive: 
\begin{enumerate}
\item An \emph{$X$-query}, represented by $(X, a, b)$, where $a, b $ are uniformly random strings in $\{0,
  1\}^n$.\footnote{We will always assume the strings $a,b$ are sent to the prover in lexicographic order.} The expected answer is two bits $\alpha,\beta\in\{-1,1\}$. 
\item A \emph{$Z$-query}, represented by $(Z, a, b)$. Same as an $X$-query, except
  with $Z$ instead of $X$.
\item An \emph{$XZ$-query}, represented by $(X, a, Z, b)$ where $a$
  and $b$ are arbitrary binary strings such that $a\wedge b = 0^n$. This type of query is used only in the energy test, and the distribution on $a$ and $b$ depends on the Hamiltonian. The expected answer is two bits $\alpha,\beta\in\{-1,1\}$.
\item A \emph{$W$-query}, represented by $(N, a, b)$ where
  $N \in \{X', Z'\}$ and $a,b$ are uniformly random strings in $\{0,1\}^n$. The expected answer is a single bit $\alpha\in\{-1,1\}$. 
\end{enumerate}

To each query is associated an intended behavior of the prover, which is specified as part of the \emph{honest strategy} given in the following definition. 

\begin{definition}\label{def:honest-strategy}
The \emph{honest strategy} for the $r$ provers consists of the following. The provers share an $(rn)$-qubit state $\ket{\psi}$ which is obtained as the qubit-by-qubit encoding, using a CSS code as described in Section~\ref{sec:stabilizer}, of the ground state $\ket{\Gamma}$ of the local Hamiltonian $H$. Each prover holds one qubit from the encoding of each qubit of $\ket{\Gamma}$. 

Upon receiving a query, any prover performs the following depending on the type of the query:
\begin{itemize}
\item $X$-query $(X,a,b)$: measure the
  compatible observables $X(a)$ and $X(b)$ on its share of the encoded
  state, and return the two outcomes. 
\item $Z$-query $(Z,a,b)$: same as $X$-query but with observables $Z(a)$ and $Z(b)$. 
\item $XZ$-query $(X,a,Z,b)$: measure the compatible observables $X(a)$
  and $Z(b)$, and return the two outcomes.
	\item $W$-query $(N,a,b)$: measure the observable $(X(a)
  + Z(b))/\sqrt{2}$ if $N = X'$, $(X(a) - Z(b))/\sqrt{2}$ if $N = Z'$, and return the outcome.
\end{itemize}
\end{definition}

The protocol is to be performed with $r$ ``physical''
provers, but we formulate all but one of the tests (the energy test) as a
two-prover test. In this case we call the two provers ``logical'' provers. A query to the two logical provers can be mapped to a query to the $r$ physical provers as follows. One of the physical provers is chosen at random to play the role of the first logical prover, called the
\emph{special prover}. The remaining $(r-1)$ physical provers together play the role of the second logical prover, called the \emph{composite prover}.\footnote{The physical provers remain isolated throughout the protocol and are never allowed to communicate; it is only for purposes of analysis that we group $(r-1)$ physical provers into a single logical prover. In particular the physical provers are never told which logical prover they are associated with, and the distribution of queries to any physical prover is the same whether it plays the role of the special or composite prover.} For a given query $Q$ to the special prover of a type among those specified above we define a \emph{complementary query} $\overline{Q}$ for the composite prover as per the following lemma.     

\begin{lemma}
  For any $X$-query or $Z$-query, there exists a complementary query $\overline{Q}$ such that
  \begin{enumerate}
  \item The query associated to each physical prover forming the composite prover in $\overline{Q}$ is of the same type as $Q$. In particular the distribution on query strings is as specified by the query type. 
  \item If all provers apply the honest strategy and provide answers $\alpha,\beta$ to $Q$ and  $\overline{\alpha},\overline{\beta}$ to $\overline{Q}$ respectively, where $\overline{\alpha}$ and $\overline{\beta}$ are each obtained as the product of the answer to the corresponding query coming from each of the physical provers making up the composite prover, it holds that $\alpha\overline{\alpha} = \beta\overline{\beta}=+1$.
  \end{enumerate}
  \label{lem:qbar}
\end{lemma}
\begin{proof}
  Both items follow from the properties of CSS codes described in Section~\ref{sec:stabilizer}. We give the proof for an
  $X$ query $(X, a, b)$. Let the index
  of the special prover be $i$, and let $S_X$ be a stabilizer
  of the code, such that $S_X$ consists only of $X$ and $I$ Paulis and has an $X$ in
  position $i$. For each physical
  prover $j \neq i$ associated with the composite prover, if the operator in
  position $j$ of $S_X$ is $X$,  prover $j$ is sent the query $(X,
  a, b)$. Otherwise, prover $j$ is sent a uniformly random $X$-query
  $(X, c, d)$.
	
		Composite answers $\overline{\alpha},\overline{\beta}$ to the complementary
    query are determined by taking the product of the answers from all
    provers who did not receive random strings; using that $S_X$ is a stabilizer of the code  ensures that item 2 is satisfied. 
	
	In the composite query, for a given choice of $S_X$ each prover
  receives a query that is either identical to the original query, or
  is a uniformly random string; since the original query is chosen at random this is also the case for each of the physical provers associated with the composite prover. This proves item 1.
\end{proof}

The complete protocol is described in Figure~\ref{fig:protocol}. It is based on four tests. In the \emph{energy test}, the verifier asks the provers to measure a randomly chosen
term in the Hamiltonian. This test is described in more detail in Section~\ref{sec:energytest}.
The remaining three tests are called the \emph{encoding tests}; together these tests form our quantum linearity test. 
The \emph{two-query linearity test}, a variant of the classical
linearity test of Blum, Luby and Rubinfeld, is designed to show the
existence of exactly linear (in a sense to be made precise in
Section~\ref{sec:linearitytest}) observables $\Xlin$ and $\Zlin$ that
are close to the provers' actual measurement operators. In the \emph{stabilizer test}, the
provers are asked to measure a random generator of the stabilizer group associated with the code, and the verifier accepts their answers if their product is $+1$. In the
\emph{anticommutation test}, a variant of the CHSH game is played 
between the verifier and the two logical provers.

\begin{figure}[H]
\rule[1ex]{16.5cm}{0.5pt}\\
Given a local Hamiltonian $H$ in XZ form, the verifier performs the following one-round interaction with $r$ provers. The probability $p\in(0,1)$ is a parameter of the protocol that can be specified freely. 
\begin{itemize}
\item Choose one of the $r$ provers uniformly at random to be the
  \emph{special prover}. The other provers form the \emph{composite
     prover}.
\item With probability $p$, perform the \emph{energy test} described in Section~\ref{sec:energytest}.
\item With probability $(1-p)/3$ each, perform one of the following three \emph{encoding tests}:
\begin{enumerate}
\item {\bf Linearity test:} The verifier chooses a basis setting $N \in
  \{X,Z\}$ and strings $a, b,  c \in  \{0, 1\}^n$ uniformly at random. 
	He sends
  the special prover $(N, a, b)$ and the composite prover either
  $\overline{(N, a, c)}$,
  $\overline{(N, b, c)}$, or $\overline{(N, a+b, c)}$, each with probability $1/3$. \\
	The verifier accepts if answers associated with the same query
  string match, and the product of the answers associated to $a,b$ and
  $a+b$ is $+1$. 
\item {\bf Anticommutation test:} The verifier chooses basis settings $N \in
  \{X,Z\}$ and  $N' \in \{X', Z'\}$ and strings $a, b,  c \in  \{0, 1\}^n$ uniformly at random. He sends the special
  prover $(N', a, b)$ and the composite prover $\overline{(X, a, c)}$
  if $N=X$, and $\overline{(Z, b, c)}$ if $N = Z$. \\
	The verifier ignores the second answer bit from each prover, and accepts or rejects according to the following rule: if
  the inner product $a \cdot b = 0\mod 2$ (i.e. the bit-wise AND $a\wedge b$ has even Hamming weight), then he automatically accepts;
  otherwise, if the two basis settings were $Z'$ and $Z$ the verifier
  accepts if the product of the provers' answers is $-1$; otherwise, he
  accepts if the product is $+1$.
\item {\bf Stabilizer test:} The verifier chooses a basis setting $N \in
  \{X, Z\}$ and three strings $a,
  b,c \in \{0, 1\}^n$ uniformly at random. He sends
  the special prover $(N,a,b)$ and the composite prover $\overline{(N, a, c)}$.
	The verifier accepts if the product of the answers associated to the
  query string $a$ is $+1$. 
\end{enumerate}
\end{itemize}
\rule[1ex]{16.5cm}{0.5pt}
\caption{The protocol}
\label{fig:protocol}
\end{figure}

%----------------------%
\section{The quantum linearity test}
\label{sec:analysis}
%----------------------%

In this section we analyze the part of the protocol described in
Section~\ref{sec:protocol} that consists of the three encoding tests. Note that these tests do not depend on the local Hamiltonian. They can thus be considered as an independent game to be played with the $r$ provers, where each test is chosen with equal probability. 

The following lemma states the main result of this section. (To
understand the notation used in the lemma it may be useful to first
read the ensuing paragraphs on modeling arbitrary strategies for the
provers in the protocol.) 

\begin{lemma}\label{lem:quantum-linearity}
Assume a strategy $(N,\ket{\psi})$ for the provers succeeds in the encoding tests with probability at least $\omega^*_{encode}-\eps$, where $\omega^*_{encode}$ is the success probability of the strategy described in Definition~\ref{def:honest-strategy}.

Then for any prover $i\in\{1,\ldots,r\}$ and $a \in \{0,1\}^n$ there exists observables $\Xlin(a)$
and $\Zlin(a)$ acting on the $i$-th prover's register\footnote{We allow extending this register by adding ancilla qubits initialized to $\ket{0}$.} such that
$$\forall a,b\in\{0,1\}^n\qquad \Xlin(a)\Xlin(b) = \Xlin(a+b)\qquad\text{and}\qquad \Zlin(a)\Zlin(b) = \Zlin(a+b),$$
and 
\begin{equation}\label{eq:x-z-anticommute}
\frac{1}{2^{2n}}\sum_{a,b\in\{0,1\}^n}\, \big\|(\Xlin(a)\Zlin(b) - (-1)^{a \cdot b} \Zlin(b)\Xlin(a)) \ket{\psi}\big\|^2 \,=\, O\big(\eps^{1/4}\big).
\end{equation}
 Moreover, if $\sx(a)$ (resp. $\sz(b)$) is the observable that prover $i$ performs when asked an $X$-query $(X,a,b)$, and the outcome $\beta$ associated to $b$ is ignored, (resp. $Z$-query $(Z,a,b)$ with the outcome $\alpha$ associated to $a$ ignored), then   
$$\frac{1}{2^n}\sum_{a\in\{0,1\}^n}\, \big\|(\Xlin(a) - \sx(a))\ket{\psi}\big\|^2 \,=\, O\big(\eps^{1/4}\big)\qquad \text{and}\qquad \frac{1}{2^n}\sum_{b\in\{0,1\}^n}\, \big\|(\Zlin(b) - \sz(b))\ket{\psi}\big\|^2 \,=\, O\big(\eps^{1/4}\big).$$
\end{lemma}

We note that the constant $\omega^*_{encode}$ is given by
\begin{equation} \omega^*_{encode} = \frac{2}{3} + \frac{1}{3}
  \omega^*_{\text{anti-com}}, \label{eq:omegaenc} \end{equation}
where $\omega^*_{\text{anti-com}} \in (0, 1)$ is specified in the proof of Lemma~\ref{lem:anti-commute}. This is because an honest strategy
passes the linearity and stabilizer tests with probability $1$, and
the anticommutation test with probability $\omega^*_{\text{anti-com}}$.

Before proceeding with the analysis of the encoding tests we introduce some notation associated with arbitrary strategies for the provers in the protocol. We  specify a strategy using the shorthand $(N,\ket{\psi})$. Here $\ket{\psi}$ denotes the $r$-partite state shared by the
provers, and $N$ the collection of POVM that the provers apply in response to the different types of queries they can be asked. Given a query $(N,a,b)$, where $N\in\{X,Z\}$ we denote by $\{N_{ab}^{\alpha\beta}\}_{\alpha,\beta}$ the two-outcome POVM that is applied by a given prover. Although these operators may differ from one prover to another it will usually not be necessary to specify explicitly the index $i\in\{1,\ldots,r\}$ associated with the prover. Instead we will only differentiate between the special prover, whose operators will be denoted $\hat{N}$, and the composite prover, for whom the resulting operator, obtained by taking the tensor product of operators applied by each of the $(r-1)$ associated physical provers, will be denoted $\overline{N}$. These operators are local to one and $(r-1)$ provers respectively, but we will usually omit tensor product signs and write them as operators acting on the whole Hilbert space, keeping in mind that an operator of the form $\hat{N}$ always commutes with an operator of the form $\overline{N}$. 

By taking appropriate marginals over the answers we can define associated observables for the provers, $\hat{X}(a)$ and $\hat{Z}(b)$ for the special prover and $\overline{X}(a)$ and $\overline{Z}(b)$ for the composite prover, where 
\begin{equation}\label{eq:hatx-def}
\hat{X}(a) = \frac{1}{2^n}\sum_{b\in\{0,1\}^n} \sum_{\beta\in \{\pm1\}} \big(N_{ab}^{1\beta}-N_{ab}^{-1\beta}\big), \qquad \hat{Z}(b) = \frac{1}{2^n}\sum_{a\in\{0,1\}^n} \sum_{\alpha\in \{\pm1\}} \big(M_{ab}^{\alpha1}-M_{a,b}^{\alpha-1}\big),
\end{equation}
for the POVM $\{N_{ab}^{\alpha\beta}\}$ and $\{M_{ab}^{\alpha\beta}\}$ associated to the queries $(X,a,b)$ and $(Z, a, b)$ respectively. Observables $\bar{X}(a)$ and $\bar{Z}(b)$ are defined similarly. 

With the notation in place the proof of Lemma~\ref{lem:quantum-linearity} follows from the analysis of the encoding tests given in the following subsections.

\begin{proof}[Proof of Lemma~\ref{lem:quantum-linearity}]
Fix an arbitrary strategy $(N,\ket{\psi})$ for the provers. For
$a,b\in\{0,1\}^n$ let $\Xlin(a)$ and $\Zlin(b)$ be the observables
introduced in Definition~\ref{def:exactly-linear}. When $a\cdot b = 1
\mod 2$ the anticommutation property implied
by~\eqref{eq:x-z-anticommute} is proven in
Lemma~\ref{lem:anti-commute}. When $a\cdot b = 0 \mod 2$ the
corresponding commutation is proved in
Lemma~\ref{lem:commutation}. Finally the relation between the
observables $\Xlin,\Zlin$ and the provers' original strategy follows
from the definition and Lemma~\ref{lem:lin-test} analyzing the
linearity test. 
\end{proof}

\subsection{Two-query linearity test}
\label{sec:linearitytest}

In~\cite{IV12} it is shown that the classical
3-query linearity test of Blum, Luby, and Rubinfeld~\cite{BLR93} (BLR) is sound against
entangled provers. The proof is an adaptation of the Fourier-analytic proof due to Hast{\aa}d to the matrix-valued setting. Here we  analyze a
two-query version of the test, again using Fourier analysis to prove
soundness. The test is based on the idea of oracularization with a dummy question
introduced in~\cite{IKM09}. We note that the use of two provers, rather than three as in
the original test, is essential for us. This is because our quantum linearity test relies on simultaneously testing for linearity of two functions that are obtained (in the honest case) by applying tensor products of $X$ and $Z$ operators respectively. For the ``linearity'' part of the test to be compatible with the other subtests, such as the anticommutation test, it is necessary that the special prover and the composite prover share a state that is equivalent to an EPR pair. Monogamy of entanglement thus prevents us from designing a protocol that would enforce both the anticommutation test and a three-prover variant of the classical linearity test; this is a key difference between our \emph{quantum} linearity test and the entangled-prover \emph{classical} linearity test of~\cite{IV12}.
% --- here we are testing for a quantum analogue of linearity instead of linearity of a classical Boolean function. 

We describe the test as a test to be performed with two provers. In the actual protocol one of the two provers is the special prover and the other is the composite prover that consists of the combination of $(r-1)$ out of the $r$ ``physical'' provers. We identify the composite prover with the first prover in the test as described below, and the special prover with the second prover in the test. When the test is performed the role of special prover is assigned uniformly at random among the $r$ possibilities.\footnote{As previously mentioned, a prover is never told if it is playing the role of the special prover or of one of the provers making up the composite prover.}

The test is specified in Figure~\ref{fig:protocol}. For convenience we repeat it here. First the verifier chooses a random basis setting $N\in\{X,Z\}$ that is sent to both provers. The accompanying strings are determined as follows:
\begin{enumerate}
\item Choose two strings $a,b \in \{0,1\}^n$ uniformly at random. Send the lexicographically ordered pair $\{a,b\}$ to the first prover.
\item Let $c$ be with equal
  probability either $a$, $b$, or $a+b$, and let $c'$ be a random
  string. Send the lexicographically ordered pair $\{c,c'\}$ to the second prover.
	\item The provers reply with $\alpha,\beta\in\{\pm 1\}$ and $\gamma,\gamma'\in\{\pm 1\}$ respectively. Depending on the value of $c$ the verifier performs one of the following two tests:
	\begin{enumerate}
\item
  \emph{Consistency test}: if $c = a$ (resp. $b$), accept if both provers return the same value as their corresponding answer: $\gamma=\alpha$ (resp. $\gamma=\beta$).
\item
  \emph{Linearity test}: if $c = a + b$, accept if $\gamma = \alpha \beta$.
\end{enumerate}
\end{enumerate}

We show the following.

\begin{lemma}\label{lem:lin-test}
Suppose two provers sharing entangled state $\ket{\psi}$ and making measurements $\{M_{ab}^{\alpha\beta}\}_{\alpha,\beta}$, $\{N_{ab}^{\alpha\beta}\}_{\alpha,\beta}$ respectively succeed in the oracularized linearity test with probability $1-\eps$. Then there exists a projective measurement $\{C^u\}_{u\in\{0,1\}^n}$ and $\eps_{lin} = O(\sqrt{\eps})$ such that  
$$\E_{a} \CON_{\psi''} (\tilde{N}_a, C_a)\,=\, 1-\eps_{lin},$$
where the expectation is taken with respect to the uniform distribution on $a\in\{0,1\}^n$, $\ket{\psi''}$ is $\ket{\psi}$ tensored with local ancilla
registers on the second prover's register, $\tilde{N}_a^{\a} =
\E_{b}\sum_{\beta} N_{ab}^{\a \beta}$ and $C_a = \sum_u (-1)^{u\cdot a}
C^u$. 

	Moreover, the honest strategy (see Definition~\ref{def:honest-strategy}) succeeds in the test with probability $1$.
\end{lemma}

Before proceeding with the proof of the lemma we introduce useful notation. 

\begin{definition}[Exactly  linear observables]\label{def:exactly-linear}
Let $(N,\ket{\psi})$ be a strategy for the provers in the protocol. For any prover $i$ and $a,b\in\{0,1\}^n$ define observables  $\Xlin(a)$ and $\Zlin(b)$ as the observables $C_a$ and $C_b$ from Lemma~\ref{lem:lin-test} when the second prover's measurements in the linearity test are  prover $i$'s measurements on an $X$-query and $Z$-query respectively. We say that $\Xlin$ and $\Zlin$ are \emph{exactly linear} observables because they automatically satisfy
\begin{equation}\label{eq:exactly-linear}
\forall a,b\in\{0,1\}^n,\qquad \Xlin(a) =  \Xlin(b)\Xlin(a+b)\qquad \text{and} \qquad \Zlin(b) =  \Zlin(a)\Zlin(a+b).
\end{equation}
\end{definition}

\begin{proof}[Proof of Lemma~\ref{lem:lin-test}]
Assume without loss of generality that the provers' POVMs
$\{M_{ab}^{\a\beta}\}_{\a,\beta}$ and $\{N_{ab}^{\alpha\beta}\}_{\alpha,\beta}$
respectively are projective. Note that here the subscript $ab$
should always be understood as a lexicographically ordered pair  $\{a,b\}$, and
the superscript indicates that $\a$ is the answer associated to $a$ and
$\beta$ to $b$.  
 Consider the
marginalized operators obtained by averaging out over the second
question: 
\begin{align*}
  \tilde{M}_a^{\a} = \E_{b} \sum_{\beta} M_{a b}^{\a \beta} ,\qquad
  \tilde{N}_a^{\a} = \E_{b} \sum_{\beta} N_{a b}^{\alpha \beta}. 
\end{align*}
These operators are in general \emph{not} projectors. It
will also be useful to consider the following conditional
measurement operator, which \emph{is} a projector:
\begin{align*}
  M_{a|a b}^{\a} &= \sum_{\beta} M_{ab}^{\a \beta}.
\end{align*}
Suppose that the provers' acceptance probability conditioned on the verifier performing the consistency part of the test (i.e. $c=a$ or $c=b$) is $1 -
\eps_{c}$, while conditioned on the verifier performing the linearity part of the test (i.e. $c=a+b$) it is $1 -
\eps_{l}$, so that $\eps = 2\eps_c/3 + \eps_l/3$. These probabilities can be translated into statements on the provers' measurements as follows.
\begin{align}
  1 - \eps_{c}  &= \E_{a} \sum_{\a} \bavg{
  \tilde{M}_a^{\a} \tilde{N}_a^{\a} }{\psi} \notag\\
  &= \E_{a} \CON_\psi(\tilde{M}_a, \tilde{N}_a) \notag\\
  &= \E_{ab} \CON_\psi(M_{a|ab}, \tilde{N}_a), \label{eq:lin-test-1a}\\
  1 - \eps_{l} &= \E_{ab} \sum_{\a, \beta} \bavg{ M_{ab}^{\a \beta}
  \tilde{N}_{a + b}^{(\a \beta)}}{\psi}\notag \\
  &=  \E_{ab} \sum_{\alpha, \beta} \bavg{ M_{a  |ab}^{\a} M_{b |ab}^{\beta} \tilde{N}_{a+b}^{(\a
   \beta)}}{\psi},\label{eq:lin-test-1b}
\end{align}
where the last equality uses that the POVM elements $M_{ab}^{\a\beta}$ are projectors. Using again Naimark's dilation theorem the POVM
$\{\tilde{N}^\a_a\}$ acting on the second register of $\ket{\psi}$ can be simulated by a
projective measurement $\{Y^\a_a\}$ acting on the state $\ket{\psi'} = \ket{\psi} \otimes
2^{-n/2} \sum_{a} \ket{a}$. 
Next we construct a set of ``exactly linear'' observables using the
projectors $Y^\a_a$. Let
$d(a | ab ) = d_{\psi'}\left(M_{a|ab }, Y_{a} \right)$, so that
by \lemref{condist},
\[ d(a | ab)^2 = O(\CON_{\psi'}(M_{a|ab}, Y_a)), \]
and taking expectation values and applying Jensen's inequality
\begin{align}
  \E_{ab} d(a|ab) &\leq \sqrt{\E_{ab} d(a | ab)^2}\notag\\
	&= O\Big(\sqrt{\E_{ab} \CON_{\psi}(M_{a|ab}, \tilde{N}_a)}\Big)\notag \\
                        &= O(\sqrt{\eps_{c}}).\label{eq:lin-test-2}
\end{align}
Introduce observables $Y_a = Y_a^{+1} - Y_a^{-1}$. For every $u\in\{0,1\}^n$ consider the Fourier transform  $\hat{Y}_u = \E_{a} (-1)^{a\cdot u} Y_a$. Define measurement operators
  $B^u = (\hat{Y}_u)^2$.
By Parseval's identity, these operators form a POVM. Using Naimark's theorem there exists an extended state $\ket{\psi''}$ and a projective
measurement $\{C^u\}$ that simulates $B^u$. Introduce the
binary projective measurement
\[ C^\a_a = \sum_{u:(-1)^{u\cdot a} = \a} C^u , \]
and the corresponding observable
\[ C_a = C_a^{+1} - C_a^{-1} = \sum_{u} (-1)^{u\cdot a} C^u. \]
From the orthogonality of the projectors $C^u$ it follows that $C_a
C_b = C_{a+b}$, so that $\{C_a\}$ is perfectly linear. It remains to
show that the operators $C_a^\a$ are consistent with the second
prover's operators $\tilde{N}_a^\a$, on the state $\ket{\psi''}$ (where
we extend $\tilde{N}_a^\a$ by making it act as the identity on the
ancilla registers). Write
\begin{align*}
  \E_{a} \CON_{\psi''} (\tilde{N}_a, C_a) &= \E_{a}\sum_\a \Re\big( 
                                          \bavg{ \tilde{N}_a^\a
                                          C_a^\a }{\psi''} \big)\\
                                        &= \E_{a}\sum_\a 
                                          \sum_{u:(-1)^{u\cdot a} = \a}
                                          \Re\big(\bavg{Y_a^\a
                                          B^u}{\psi'}\big) \\
                                        &= \E_{a}\sum_\a 
                                          \sum_{u:(-1)^{u\cdot a} = \a}
                                          \Bavg{\frac{1}{2}\big(1 +
                                          (-1)^\a Y_a\big) (\hat{Y}_u)^2
                                          }{\psi'} \\
                                        &= \frac{1}{2} + \frac{1}{2}
                                          \E_a \sum_u (-1)^{u\cdot a}
                                          \bavg{ Y_a
                                          (\hat{Y}_u)^2 }{\psi'} \\
                                        &= \frac{1}{2} + \frac{1}{2}
                                          \E_a \bavg{\hat{Y}_u^3
                                          }{\psi'} .
																					\end{align*}
To conclude, this last expression can be bounded as
\begin{align*}
  \sum_u \bavg{\hat{Y}_u^3}{\psi'} &= \sum_u \bavg{(\E_{abc} (-1)^{u \cdot (a + b +
                                 c)} Y_{a} Y_{b} Y_{c})}{\psi'} \\
            &= \E_{ab} \avg{(Y_{a} Y_{b} Y_{a + b})}{\psi'} \\
            &= \E_{ab} \sum_{\a \beta} \bavg{(Y_{a}^{\a} Y_{b}^{\beta} Y_{a+b}^{\a \beta} - Y_{a}^{\a} Y_{b}^{\beta} Y_{a
              + b}^{- \a \beta})}{\psi'} \\
            &= \E_{ab} \sum_{\a_1 \a_2}
              \bavg{(2Y_{a}^{\a} Y_{b}^{\beta} Y^{\a \beta}_{a
              + b} - Y_{a}^{\a} Y_{b}^{\beta})}{\psi'} \\
            &= 2\E_{ab} \sum_{\a \beta} \bavg{(Y_{a}^{\a} Y_{b}^{\beta} Y^{\a \beta}_{a
              + b})}{\psi'} - 1 \\
            &\geq 2\E_{a b} \Big( \sum_{\a \beta}
              \bavg{(M_{a|ab}^{\a} M_{b|ab}^{\beta}
              Y^{\a \beta}_{a + b})}{\psi'} - O\big( d(a|ab) +
              d(b|ab)\big)\Big)  - 1\\
            &= 1 - O\big(\eps_{l}+ \sqrt{\eps_{c}}\big),
\end{align*}
where the inequality uses Lemma~\ref{lem:approx} twice and the last line is by~\eqref{eq:lin-test-1b} and~\eqref{eq:lin-test-2}.

\end{proof}

\subsection{Stabilizer Test}

 %We pick a random basis setting $N \in \{X, Z\}$ and
 %random strings $a, b, c \in \{0, 1\}^n$, and send the special prover the
 %query $(N, a, b)$ and the composite prover the query $\overline{(N,a,c)}$. We
 %discard the second answer bit from each prover, and accept if the
 %product of the first answer from the special player and the composite
 %player is equal to $+1$.

The stabilizer test is described in Figure~\ref{fig:protocol}. The following lemma states the main consequence we will use. 

\begin{lemma}\label{lem:stabilizer}
Suppose the strategy $(N,\ket{\psi})$ succeeds in the stabilizer test with probability $1-\eps$. Then there exists $\eps_{stab} = O(\sqrt{\eps})$ such that 
\[ \Big(\E_a \big\|(\sx(a) - \cx(a) )\ket{\psi} \big\|^2\Big)^{1/2}  \leq \eps_{stab} \qquad \text{and} \qquad  \Big(\E_a \big\|(\sz(a) - \cz(a)) \ket{\psi} \big\|^2\Big)^{1/2} \leq \eps_{stab}, \]
where $\sx(a)$, $\sz(b)$ and $\cx(a)$, $\cz(b)$ are observables defined from the provers' strategies in~\eqref{eq:hatx-def}.

	Moreover, the honest strategy succeeds in the test with probability $1$.
\end{lemma}
\begin{proof}
It follows from the definition of $\CON_\psi$ that any strategy $(N,\ket{\psi})$ succeeding in the test with probability $1 - \eps$ satisfies 
 \begin{align*}
   \E_{a} \CON_\psi(\sx(a), \cx(a)) &\geq 1 - O(\eps) \\
   \E_{a} \CON_\psi(\sz(a), \cz(a)) &\geq 1 - O(\eps).
 \end{align*}
 By applying Lemma~\ref{lem:condist} to the above relations, we obtain:
$$\E_{a} d_\psi(\sx(a), \cx(a)) = O(\sqrt{\eps})\qquad \text{and}\qquad \E_{a} d_\psi(\sz(a), \cz(a)) = O(\sqrt{\eps}),$$
Expression~\eqref{eq:dist_observables} for the state-dependent distance $d_\Psi$ between two observables yields a bound on the squared norm.
\end{proof}

\subsection{Anticommutation Test}

The anticommutation test is described in Figure~\ref{fig:protocol}. The goal of the test is to certify that 
$$\sx(a)\sz(b)\ket{\psi}
\approx (-1)^{a \cdot b} \sz(a)\sx(b) \ket{\psi},$$
 where $\sx(a)$ and $\sz(b)$ are the observables defined from the provers' strategies in~\eqref{eq:hatx-def}.
 There are two cases: if
$a$ and $b$ overlap on an even number of positions, then the two operators should commute;
otherwise, they should anti-commute. The anticommutation test enforces the latter property. In Section~\ref{sec:commutation} we show how the former can be derived as a consequence.

\begin{lemma}\label{lem:anti-commute}
There exists $\omega^*_{\text{anti-com}} \in (0,1)$ such that the following holds. Suppose the strategy $(N,\ket{\psi})$ succeeds in the
anticommutation test with probability $\omega_{\text{anti-com}}^*-\eps$ and in
the stabilizer test with probability $1 - \eps_{stab}$. Then there exists
$\eps_{ac} = O(\sqrt{\eps}) + O(\sqrt{\eps_{stab}})$ such that 
$$\Big(\E_{a, b : a\cdot b = 1} \big\| \big(\sx(a)\sz(b) -
(-1)^{a \cdot b} \sz(b)\sx(a) \big)\ket{\psi}\big\|^2\Big)^{1/2}
= \eps_{ac}.$$
Moreover, the honest strategy succeeds in this test with
probability $\omega^*_{\text{anti-com}}$.
\end{lemma}
We remark that the constant $\omega_{\text{anti-com}}^*$ is
closely related to the maximum quantum winning probability
of the CHSH game.
\begin{proof}
The analysis follows very closely that of the ``special-player''
stabilizer game in~\cite[Section~3.2]{ji2015classical}, which is in
turn based on the CHSH game. The key idea is to show the
existence of ``rotated'' operators $\sx'$ and $\sz'$ acting on the special
prover such that $\frac{\sx' + \sz'}{\sqrt{2}}$ (resp. $\frac{\sx'
  - \sz'}{\sqrt{2}}$) is consistent with $\cx(a)$
(resp. $\cz(b)$) on the composite prover. Since we know from the analysis of the
stabilizer test (Lemma~\ref{lem:stabilizer}) that $\cx(a)$ and $\cz(b)$ are consistent with
$\sx(a)$ and $\sx(b)$ respectively, this
allows us to conclude that $\frac{\sx' + \sx'}{\sqrt{2}}$ is consistent
with $\sx(a)$ and $\frac{\sx' - \sz'}{\sqrt{2}}$ is consistent with
$\sz(b)$. Since the two operators $\frac{\sx' \pm \sz'}{\sqrt{2}}$
anticommute \emph{exactly} with each other by construction, the
operators $\sx(a)$ and $\sz(b)$ must thus \emph{approximately}
anti-commute. Note that the rotated operators $\sx'$ and $\sz'$ are allowed
to depend on both $a$ and $b$; we leave this dependence implicit in the notation and similarly suppress it from $\cx(a)$ and $\cz(b)$.

We now analyze the test in more detail. First, recall that as stated
in the protocol, $a$ and $b$ are chosen independently and uniformly at
random, so they have a chance of $1/2$ of having inner product $0$,
and in this case the test always accepts. Let $p_{suc}$ be the success probability of the
test and define the bias $\beta$ as: 
\begin{align*}
  \beta = 16p_{suc}-12 &= 8 \E_{a,b: a \cdot b = 1} \left(\frac{1}{2} + \frac{1}{8} \langle  \sx' \cx + \sz' \cx + \sx' \cz
- \sz' \cz \rangle_\Psi \right) -4 \\
  &= \langle  \sx' \cx + \sz' \cx + \sx' \cz
- \sz' \cz \rangle_\Psi.
\end{align*}
The bias can be decomposed as a sum of squares as follows:
\[
  \frac{2\beta}{\sqrt{2}} = \E_{a,b: a \cdot b = 1} \Big( 4 - \Big\langle\Big(\frac{\sx' + \sz'}{\sqrt{2}} - \cx \Big)^2\Big\rangle_{\Psi} - \Big\langle\Big(\frac{\sx' - \sz'}{\sqrt{2}} - \cz \Big)^2\Big\rangle_{\Psi}\Big).
\]
Let $\omega^*_{\text{anti-com}} = \frac{3}{4} + \frac{\sqrt{2}}{8}$. If the success
probability is $p_{suc} = \omega^*_{\text{anti-com}} - \eps$, then $\beta =
2\sqrt{2} - 16 \eps$. From this and the sum-of-squares decomposition
above, we deduce
\begin{align}\label{eq:sos-1}
  \E_{a,b: a \cdot b = 1} \Big\langle\Big(\frac{\sx' + \sz'}{\sqrt{2}} - \cx \Big)^2\Big\rangle_{\Psi}\leq
                                                                        \frac{32}{\sqrt{2}}
                                                                        \eps ,\qquad
  \E_{a,b: a \cdot b = 1} \Big\langle\Big(\frac{\sx' - \sz'}{\sqrt{2}} - \cz \Big)^2\Big\rangle_{\Psi} \leq
                                                                        \frac{32}{\sqrt{2}}
                                                                        \eps.
\end{align}
%Or equivalently,
%\begin{align*}
%  \E_{a,b} \left\| \left(\frac{\sx' + \sz'}{\sqrt{2}} - \cx(a) \right)
%  \ket{\psi}\right\| \leq
%  O(\sqrt{\eps}) ,\qquad
%  \E_{a,b} \left\| \left(\frac{\sx' - \sz'}{\sqrt{2}} - \cz(b)\right) \ket{\psi} \right\| \leq
%  O(\sqrt{\eps}).
%\end{align*}
  Now, for \emph{any} observables $\sx', \sz'$, the following anti-commutation relation holds:
  \begin{align*}
    \{ \sx' + \sz', \sx' - \sz' \} &= (\sx' + \sz')(\sx' - \sz')+ (\sx' - \sz')(\sx' + \sz') \\
                         &= I - \sx'\sz' + \sz' \sx' - I + I + \sx' \sz' - \sz' \sx' -
      I \\
    &= 0.
  \end{align*}
  Thus, the operators $(\sx' + \sz')/\sqrt{2}, (\sx' - \sz')/\sqrt{2}$ are
  exactly anticommuting. The anti-commutator of $\sx(a)$ and $\sz(b)$ is
  \begin{align*}
    \E_{a,b: a \cdot b = 1}& \big\|  (\sx(a) \sz(b) - \sz(b) \sx(a)) \ket{\psi}\big\|^2 \\
    &= \E_{a,b: a \cdot b = 1} \big\|(\cx(a) \cz(b)
      -\cz(b)\cx(a))\ket{\psi}\big\|^2 +
                                                          O(\sqrt{\eps_{stab}}) \\
    &= \frac{1}{4}
      \E_{a,b: a \cdot b = 1} \big\| ((\sx' +
      \sz')(\sx' - \sz') - (\sx'
      - \sz')(\sx' +
      \sz'))\ket{\psi}
      \big\|^2 +
      O(\sqrt{\eps}) + O(\sqrt{\eps_{stab}}) \\
    &= O(\sqrt{\eps}) + O(\sqrt{\eps_{stab}}),
  \end{align*}
  where we replaced the $X$ and $Z$ operators first by $\cx,
  \cz$, and then by $(\sx' \pm \sz') /
  \sqrt{2}$, and bounded the error using~\eqref{eq:sos-1} and the Cauchy-Schwarz inequality. This proves the lemma.
\end{proof}

\subsection{Commutation}\label{sec:commutation}

The protocol does not involve a test for commutation, as the required property can be derived as a consequence of the encoding tests. 

\begin{lemma}\label{lem:commutation}
Suppose the strategy $(N,\ket{\psi})$ succeeds in the anti-commutation, linearity and stabilizer tests with probability $1-\eps$ each. Then there exists $\eps_{com} = O(\sqrt{\eps})$ such that 
$$\Big(\E_{a,b:a\cdot b=0} \big\| \big(\sx(a)\sz(b) -  \sz(b)\sx(a) \big)\ket{\psi}\big\|^2\Big)^{1/2} = \eps_{com}.$$ 
\end{lemma}

\begin{proof}
We combine the anticommutation, linearity, and stabilizer tests through the following sequence of approximate identities, where the symbol $\approx$ means that two states are
$O(\eps_{lin} + \eps_{stab} + \eps_{anti-com})$-close in squared Euclidean
norm. In particular, we use the stabilizer test to ``push'' an operator
from the special prover onto the other provers, which allows us to
commute it past other operators acting on the special prover. First we relate the observables associated with the special prover's strategy to the exactly linear observables $\Xlin$, $\Zlin$ obtained from the linearity test (see Definition~\ref{def:exactly-linear}). 
\begin{align*}
\E_{a,b: a \cdot b = 0} \sx(a) \sz(b) \ket{\psi} &\approx \E_{a,b: a \cdot b = 0} \sx(a) \cz(b) \ket{\psi}\\
&\approx \E_{a,b: a \cdot b = 0} \Xlin(a) \cz(b)\ket{\psi}\\
&\approx \E_{a,b: a \cdot b = 0} \Xlin(a) \sz(b)\ket{\psi}\\
&\approx \E_{a, b: a \cdot b =0}  \Xlin(a) \Zlin(b)\ket{\psi},
\end{align*}
where the first and third lines follow from the stabilizer test (Lemma~\ref{lem:stabilizer}) and the second and fourth from the linearity test (Lemma~\ref{lem:lin-test}). Next write
\begin{align*}
\E_{a, b: a \cdot b =0}  \Xlin(a) \Zlin(b)\ket{\psi}   &= \E_{a,b : a \cdot b = 0}
                                             \E_{c: c\cdot a= 1} \Xlin(a)
                                             \Zlin(c) \Zlin(c + b)
                                             \ket{\psi} \\
                                        &\approx \E_{a,b : a \cdot b = 0}
                                          \E_{c: c\cdot a= 1} \cz(c+b) \Xlin(a) \Zlin(c) \ket{\psi} \\
                                        &\approx -\E_{a,b : a \cdot b = 0} \E_{c: c\cdot a= 1} \cz(c +
                                          b) \Zlin(c) \Xlin(a) \ket{\psi} \\
                                        &\approx -\E_{a,b : a \cdot b = 0} \E_{c: c\cdot a = 1} \Zlin(c) \Xlin(a)
                                          \Zlin(c+ b) \ket{\psi} \\
                                        &\approx \E_{a,b : a \cdot b =0} \E_{c: c\cdot a = 1} \Zlin(c) \Zlin(c+b)
                                          \Xlin(a) \ket{\psi} \\
                                        &= \E_{a,b: a\cdot b =
                                          0} \Zlin(b) \Xlin(a)
                                          \ket{\psi} \\
                                        &\approx \E_{a,b: a\cdot b =
                                          0} \sz(b) \sx(a) \ket{\psi}.
\end{align*}
Here the first equality uses the exact linearity~\eqref{eq:exactly-linear} of $\Xlin$ and
$\Zlin$. The second line uses the linearity test and the stabilizer test. 
The third line uses approximate anticommutation (Lemma~\ref{lem:anti-commute}). The fourth line again uses the stabilizer and linearity tests, and the fifth line uses approximate anticommutation. The sixth line uses exact linearity, and the last is obtained from the linearity and stabilizer tests.
\end{proof}

%================================================%
\section{A game for the local Hamiltonian problem}
\label{sec:game-lh}
%================================================%

In this section we complete the analysis of the protocol described in Section~\ref{sec:protocol} and prove our main theorem. The encoding tests have already been considered in the previous section, and it remains to define and analyze the energy test; this is done in the following subsection. In Section~\ref{sec:isometry} we introduce an isometry that will let us ``extract'' an $n$-qubit state, destined to play the role of the $\QMA$ witness for the local Hamiltonian instance under consideration, from the strategy of the provers. Theorem~\ref{thm:main} in Section~\ref{sec:main-proof} summarizes the result of the analysis so far, stating it in terms of a gap-preserving reduction. Finally Theorem~\ref{thm:main0} is proved in Section~\ref{sec:amplification} by combining Theorem~\ref{thm:main} with a gap amplification procedure.

\subsection{Energy Test}
\label{sec:energytest}

In the energy test the verifier asks the provers to measure a local
Hamiltonian on their shared encoded state. The test is made of two subtests, each to be performed with probability half: the \emph{energy measurement test}  and the \emph{consistency test}. 

\subsubsection{Energy measurement test}

The goal of the energy measurement test is to estimate the energy of a randomly
chosen  term in the Hamiltonian. We analyze the test for the general case of a 
Hamiltonian that is not necessarily local, but such that $H$ can be
decomposed as 
\begin{equation}\label{eq:xz-hamiltonian}
 H = \frac{1}{m} \sum_{\ell = 1}^{m} \alpha_\ell P_\ell,
\end{equation}
where each $P_\ell$ is an $n$-qubit operator consisting of a tensor
product of single-qubit $I$, $X$ and $Z$ Pauli operators and the real coefficients $\alpha_\ell$
satisfy $|\alpha_\ell| \leq 1$ for all $\ell\in\{1,\ldots,m\}$. The energy measurement test proceeds as follows:
\begin{itemize}
\item For each $\ell\in\{1,\ldots,m\}$ define an operator $Q_\ell$ acting on $rn$ qubits by
  replacing each Pauli $X$ in $P_\ell$ with $X_{logical}$ on the $r$-qubit code
  state, and each Pauli $Z$ by $Z_{logical}$.
\item Send each prover an $XZ$-query (see Section~\ref{sec:protocol}) representing the associated share of $Q_\ell$.
\item Each prover replies with two values in $\{-1,1\}$. Take the product of all values received, and compare it to the sign of $\alpha_\ell$. If the
  signs disagree, accept. If the signs
 agree, reject with probability $|\alpha_\ell|$ and accept with probability $1-|\alpha_\ell|$.
\end{itemize}

\begin{lemma}
  The acceptance probability of the energy measurement test, when the correct Pauli operators are applied by each prover on its respective register of an $(rn)$-qubit state $\ket{\psi}$, is
\begin{align*}
w^*_{energy}(H,\ket{\psi}) &= 1-\Big( \frac{1}{2m} \sum_{\ell = 1}^{m} \frac{ |\alpha_{\ell} | +
  \alpha_{\ell} \langle \psi|P_\ell |\psi \rangle}{2} \Big)\\
  &= 1-\Big(\frac{1}{4}\bra{\psi} H\ket{\psi} + \frac{1}{2m }
    \sum_{\ell} |\alpha_{\ell}|\Big).
\end{align*}
\label{lem:energy_test}
\end{lemma}

\begin{proof}
The proof is a simple calculation in all points similar to that performed in~\cite[Section~4]{ji2015classical}; see in particular the discussion that precedes Theorem~23 in that paper. We omit the details.
\end{proof}
Note that this lemma only describes the behavior of the honest
provers. The corresponding soundness result for dishonest provers is
essentially our main theorem (Theorem~\ref{thm:main}).

\subsubsection{Consistency test}

The goal of the consistency test is to ensure that the measurement operators used by the provers in the energy test are consistent with the operators $\Xlin$ and $\Zlin$ defined from their strategies in~\eqref{eq:hatx-def}. In the test the verifier first selects a local term $P_\ell $ from the Hamiltonian uniformly at random. The energy measurement test considers an associated term $P_\ell^i$, $i=1,\ldots,r$, for each prover, which can be written as a product $X(a_i)Z(b_i)$ of $X$ and $Z$ operators for non-overlapping strings $a_i$ and $b_i$. Let $i$ be the index of the special prover. The verifier performs one of the following tests, each chosen with the indicated probability.
\begin{itemize}
\item With probability $1/2$, send the special prover $(X,a,Z,b)$, and the composite prover $\overline{(X, c, c + a)}$, where $c \in \{0, 1\}^n$ is chosen uniformly at random. Accept if the special prover's $X$-answer agrees with the product of the composite prover's two answers.
\item With probability $1/4$, send the special prover $(X, c, d)$, and the composite prover $\overline{(X, c, c + a)}$, where $c, d \in \{0, 1\}^n$ are chosen uniformly at random. Accept if the special prover and composite prover agree on $X(c)$.
\item With probability $1/4$, send the special prover $(X, c+a,d)$, and the composite prover $\overline{(X, c, c + a)}$, where $c, d \in \{0, 1\}^n$ are chosen uniformly at random. Accept if the special prover and composite prover agree on $X(c+a)$.
% \item With probability $1/6$, send the special prover $(X, c, c+a)$ and the composite prover $\overline{(X, c, c+a)}$, where $c \in \{0, 1\}^n$ is chosen uniformly at random. Accept if the product of the answers for $c$ and the product of the answers for $c+a$ are both $+1$.
\end{itemize}
The same tests are performed with the role of $X$ and $Z$ (to the composite prover) interchanged, the $X$ and $Z$-variants being selected by the verifier with probability $1/2$ each. 

\begin{lemma}
  Suppose the strategy $(N,\ket{\psi})$ for the provers succeeds in the consistency test with probability $1 -
  \eps_{cons}$ and in the encoding tests with probability $1 - \eps_{encode}$, then
  \[\frac{1}{m}\sum_{\ell=1}^m \big
	\| (\sp_\ell^i - \Xlin(a)\Zlin(b))\ket{\psi}\big\|^2 \,=\,
  O\big(\eps_{cons}^{1/4}\big) + O\big(\eps_{encode}^{1/4}\big), \]
where $a$ and $b$ are strings such that $P_\ell^i = X(a)Z(b)$, and $\sp_\ell^i$ is the measurement applied by the special prover upon receiving the query $(X, a, Z, b)$ in the energy test. 
 
Moreover, honest provers (see Definition~\ref{def:honest-strategy}) pass the test with probability $1$.
\label{lem:energy_consistency}
\end{lemma}

\begin{proof}
  We show that $XZ$-queries, $X$-queries, and $Z$-queries
  on the special prover are all consistent with $\overline{(X, c, c + a)}$ and
  $\overline{(Z, c, c+b)}$  queries to the composite prover. The analysis uses similar techniques to the analysis of the linearity test. First, let us
  analyze the $X$ case. Let the POVM applied by the composite prover
  be $\{M_{c, c+a}^{\alpha \alpha'}\}$ and define marginalized
  operators 
  \[    M_{c | c, c+a}^{\alpha} = \sum_{\alpha'} M_{c,
    c+a}^{\alpha\alpha'}.\]
Likewise, define marginalized operators for the special prover:
\[ \sp^{\alpha}_{a|\ell} = \sum_{\alpha'} P^{\alpha \alpha'}_{\ell}, \qquad \sp^{\alpha'}_{b | \ell} = \sum_{\alpha} P^{\alpha \alpha'}_{\ell}. \]
Here we have suppressed the superscript $i$ indicating the index of the special prover. 

 The following consistency relations follow from the assumption that the provers succeed with probability $1-\eps_{cons}$ in the test. We use the notation $\E_{a \sim P_\ell}$ to indicates that the string $a$ is chosen from the distribution of queries induced by the Hamiltonian, in contrast to $\E_a$ which indicates a uniformly random string.
	  \begin{align}
      \E_{a \sim P_\ell} \E_{c} \CON(M_{c | c, c+a}^\alpha, \sx(c)^\alpha) &\geq 1 -
                                                                        O(\eps_{cons}) \label{eq:energy_cons1}\\
      \E_{a \sim P_\ell} \E_{c} \CON(M_{c+a | c, c+a}^\alpha, \sx(c + a)^\alpha) &\geq 1 -
                                                                                O(\eps_{cons}) \label{eq:energy_cons2}\\
      \E_{a \sim P_\ell} \E_c \CON(\sp_{a | \ell}^{\alpha}, \sum_{\beta \cdot \beta' =
      \alpha}M_{c, c+a}^{\beta \beta'}) &\geq 1- O(\eps_{cons}) \label{eq:energy_cons3}.
      % \E_{a \sim P_\ell} \E_c \CON(\sx(c)^{\alpha}, \cx(c)^{\alpha}) &\geq 1 - O(\eps_{cons}) \label{eq:energy_cons4} \\
      % \E_{a \sim P_\ell} \E_c \CON(\sx(c+a)^{\alpha}, \cx(c+a)^{\alpha}) &\geq 1 - O(\eps_{cons}) \label{eq:energy_cons5}.
  \end{align}
  We now use these relations to show that the special prover's marginalized measurement $\sp^{\alpha}_{a|\ell}$ is close to the operator $\Xlin^{\alpha}(a)$ produced by the linearity test. We show this in two steps. First, we relate $\sp^{\alpha}_{a|\ell}$ to the composite prover's measurement:
  \begin{align*}
    \E_{a \sim P_\ell} &\CON\Big(\sp_{a|\ell}^{\alpha}, \Xlin^{\alpha}(a) \Big) \\
                         &\geq     \E_{a \sim P_\ell}   \E_c \Big[ \CON\Big(\sum_{\beta \cdot \beta' = \alpha} M^{\beta \beta'}_{c, c + \alpha}, \Xlin^\alpha(a) \Big)  - d_\Psi \Big(\sp^\alpha_{a | \ell}, \sum_{\beta \cdot \beta' = \alpha} M^{\beta \beta'}_{c, c+ a}\Big) \Big] \\
    &\geq \E_{a \sim P_\ell} \E_c   \CON\Big(\sum_{\beta \cdot \beta' = \alpha} M^{\beta \beta'}_{c, c + a}, \Xlin^\alpha(a)\Big) - O(\sqrt{\eps_{cons}}) \\
    &= \E_{a\sim P_\ell} \E_c \sum_{\alpha} \sum_{\beta \cdot \beta' = \alpha} \bavg{M^{\beta \beta'}_{c, c+a} \Xlin^\alpha(a)}{\Psi} - O(\sqrt{\eps_{cons}}).
\end{align*}
  In the above, we used Lemma~\ref{lem:approx} to go from the first to the second line, and then lemma~\ref{lem:condist} and~\eqref{eq:energy_cons3} to go to the third line.
Next we relate $M$ to a product of two measurements $\sx$:
\begin{align*}
    \E_{a \sim P_\ell} \CON\Big(\sp_{a|\ell}^{\alpha}, \Xlin^{\alpha}(a) \Big)  &= \E_{a\sim P_\ell} \E_c \sum_{\alpha} \sum_{\beta \cdot \beta' = \alpha} \bavg{M^{\beta}_{c| c, c+a} M^{\beta'}_{c+a|c, c+ a} \Xlin^\alpha(a)}{\Psi}  - O(\sqrt{\eps_{cons}})\\
  &\geq \E_{a\sim P_\ell} \E_c \sum_{\alpha} \sum_{\beta \cdot \beta' = \alpha} \bavg{\sx^{\beta}(c) \sx^{\beta'}(c+a) \Xlin^\alpha(a)}{\Psi} - O(\sqrt{\eps_{cons}}).
\end{align*}
Here, we used equations~\eqref{eq:energy_cons1} and~\eqref{eq:energy_cons2}, together with Lemmas~\ref{lem:approx} and~\ref{lem:condist}.
Finally, we use the encoding test to relate $\sx$ to the exactly linear observable $\Xlin$: 
\begin{align*}
  \E_{a \sim P_\ell} &\CON\Big(\sp_{a|\ell}^{\alpha}, \Xlin^{\alpha}(a) \Big) \\
	&\geq \E_{a\sim P_\ell} \E_c  \sum_{\alpha} \sum_{\beta \cdot \beta' = \alpha}  \bavg{\cx^{\beta}(c) \sx^{\beta'}(c+a) \Xlin^\alpha(a)}{\Psi}  -  O(\sqrt{\eps_{cons}} + \sqrt{\eps_{encode}})\\
	&\geq \E_{a\sim P_\ell} \E_c \Big[ \sum_{\alpha} \sum_{\beta \cdot \beta' = \alpha} \bavg{\cx^{\beta}(c)  \Xlin^{\beta'}(c+a) \Xlin^\alpha(a)}{\Psi}\\
	&\qquad\qquad - d_\Psi(\sx(c+a), \Xlin(c+a)) \Big] - O(\sqrt{\eps_{cons}}+\sqrt{\eps_{encode}}) \\
  &\geq \E_{a\sim P_\ell} \E_c \Big[ \sum_{\alpha} \sum_{\beta \cdot \beta' = \alpha} \bavg{\Xlin^{\beta}(c) \Xlin^{\beta'}(c+a) \Xlin^{\alpha}(a)}{\Psi}- d_\Psi(\sx(c), \Xlin(c)) \Big] - O(\sqrt{\eps_{cons}}+\sqrt{\eps_{encode}}) \\
  &\geq \E_{a\sim P_\ell} \E_c  \sum_{\alpha} \sum_{\beta \cdot \beta' = \alpha} \bavg{\Xlin^{\beta}(c) \Xlin^{\beta'}(c+a) \Xlin^\alpha(a)}{\Psi}  - O(\sqrt{\eps_{cons}}+\sqrt{\eps_{encode}}) \\
  &= 1 - O(\sqrt{\eps_{cons}}+\sqrt{\eps_{encode}}).
\end{align*}
Here we used Lemma~\ref{lem:stabilizer} for the first inequality, Lemma~\ref{lem:approx}  for the second, Lemma~\ref{lem:lin-test} for the third, again Lemma~\ref{lem:stabilizer} and~Lemma~\ref{lem:approx} for the third and Lemma~\ref{lem:lin-test} for the fourth. The last equality follows by exact linearity (Definition~\ref{def:exactly-linear}). Note that we could use Lemma~\ref{lem:stabilizer} here because the marginal distributions of $c$ and $c+a$ are both completely uniform.

Performing an analogous analysis for the $Z$ operators,
  \[ \E_{b \sim P_\ell} \CON(\sp_{b|\ell}^{\beta}, \Zlin(b)^\beta) \geq 1 - O(\sqrt{\eps_{cons}}) - O(\sqrt{\eps_{encode}}). \]
  Thus the special prover's operators are close to $\Xlin(a)$ and $\Zlin(b)$ used in the encoding tests. To put these results together it remains to apply the stabilizer property to these operators; while we cannot do this directly since $a$ and $b$ are not distributed uniformly, we can use the exact linearity to write $\Zlin(b) = \E_{c} \Zlin(b+c) \Zlin(c)$, and apply Lemma~\ref{lem:stabilizer} to each term in the product:
  \begin{align*}
    \E_\ell \sp_\ell\ket{\psi} &= \E_\ell \sp_{a|\ell} \sp_{b|\ell} \ket{\psi} \\
    &\approx \E_\ell \sp_{a|\ell} \Zlin(b)\ket{\psi} \\
    &= \E_{\ell} \E_{c} \sp_{a|\ell} \Zlin(b + c) \Zlin(c) \ket{\psi} \\
    &\approx \E_{\ell} \E_{c} \cz(c) \cz(b+c) \sp_{a|\ell} \ket{\psi} \\
    &\approx \E_\ell \E_{c} \cz(c)\cz(b+c) \Xlin(a) \ket{\psi} \\
    &\approx \E_\ell \E_{c} \Xlin(a) \Zlin(b+c)\Zlin(c) \ket{\psi} \\
    &= \E_{\ell} \Xlin(a) \Zlin(b) \ket{\psi}
  \end{align*}
  where the approximate equalities are measured using the distance $d_\psi$ and we consider approximations of order $O(\eps_{cons}^{1/4}) + O(\eps_{encode}^{1/4})$ (the exponent $1/4$ arises when converting $\CON$ to $d_\psi$ via lemma~\ref{lem:condist}).
\end{proof}

%-----------------------%
\subsection{The isometry}
\label{sec:isometry}
%-----------------------%

Suppose given a strategy $(N,\ket{\psi})$ for the provers that succeeds with probability $1-\eps$ in the protocol described in Section~\ref{sec:protocol}. We define an isometry that maps $\ket{\psi}$ to an $n$-qubit state $\ket{\varphi}$ such that $\ket{\varphi}$ would succeed in the energy test with a probability that deviates from the actual provers' by $O(\eps^{1/16})$, if it was measured according to the correct Pauli operators associated with the local term that the verifier chooses in the test (for any possible such choice). 

Our construction uses an isometry introduced by McKague~\cite{McKague13}, applied independently to each prover's register of $\ket{\psi}$. This isometry produces an $(rn)$-qubit state $\ket{\varphi_1}$, where $n$ qubits are defined locally for each one of the $r$ provers. The single-prover case is  described and analyzed first, in Section~\ref{sec:mckague}; the full isometry acting on all $r$ provers is described in Section~\ref{sec:full-iso}. In particular, we show in that section that the expectation value of the \emph{encoded} Hamiltonian acting on the output state of the isometry is close to the measured expectation value of the provers.

\subsubsection{The single-prover isometry}\label{sec:mckague}

We describe the isometry associated with the first provers' register; the isometries for the  remaining provers are similar. The isometry appends an $n$-qubit maximally
entangled state to $\ket{\psi}$, and ``swaps'' part of $\ket{\psi}$ from the first register to the first half of the maximally entangled state. The output
of the circuit is given by 
\[ \ket{\varphi_0} = \Phi^1(\ket{\psi}) = \frac{1}{2^{3n/2}} \sum_{z,y,w\in\{0,1\}^n} (-1)^{y \cdot (z
  +w)} \Xlin(w) \Zlin(y) \Xlin(z) \ket{\psi} \otimes \ket{wz}, \]
	where the last two registers correspond to the maximally entangled state used as ancilla. 
Here $\Xlin$ and $\Zlin$ are the exactly linear operators
 obtained from the special provers' measurement operators on $X$ and $Z$ queries respectively (see Definition~\ref{def:exactly-linear}). 

%The state $\ket{\varphi_{1}}$ is obtained by applying $\Phi$ independently to each of the provers' registers of $\ket{\psi}$ (and a separate $n$-qubit maximally entangled state for each prover). 

\begin{lemma}
For any $a, b\in\{0,1\}^n$ it holds that 
$$ \big|\bra{\varphi_0} X(a) Z(b) \ket{\varphi_0} - \bra{\psi}
\Xlin(a)\Zlin(b) \ket{\psi}\big| \,=\,O\big(\sqrt{\eps_{lin}} + \sqrt{\eps_{ac}}+\sqrt{\eps_{com}}+ \sqrt{\eps_{stab}}\big),$$
where $X(a)Z(b)$ is the corresponding tensor product Pauli operator acting on the
second register of $\ket{\varphi_0}$ (associated with the first half
of the maximally entangled state used as ancilla), $\Xlin(a), \Zlin(b)$
are the special prover's exactly linear operators (as defined in Definition~\ref{def:exactly-linear}), and $\eps_{lin}$, $\eps_{ac}$, $\eps_{com}$ and $\eps_{stab}$ are defined
in Lemma~\ref{lem:lin-test}, Lemma~\ref{lem:anti-commute},
Lemma~\ref{lem:commutation} and Lemma~\ref{lem:stabilizer}
respectively.
\label{lem:iso-single-prover-xz}
\end{lemma}

\begin{proof}
Expand 
 \begin{align*}
   \bra{\varphi_0} X(a) Z(b) \ket{\varphi_0} &= \frac{1}{2^{3n}} \sum_{yzw}
   \sum_{y'z'w'} \bra{\psi}\bra{wz}  (-1)^{y\cdot(z + w)} \Xlin(z)
   \Zlin(y) \Xlin(w) \\
   &\qquad\qquad\qquad\qquad \cdot X_2(a) Z_2(b) (-1)^{y'\cdot(z' + w')} \Xlin(w')
   \Zlin(y') \Xlin(z')  \ket{\psi}\ket{w'z'} \\
   &= \frac{1}{2^{3n}} \sum_{yzw}
   \sum_{y'z'w'} \bra{\psi}\bra{wz}  (-1)^{y\cdot(z + w)} \Xlin(z)
   \Zlin(y) \Xlin(w) \\
   &\qquad\qquad\qquad\qquad \cdot (-1)^{w'\cdot b} (-1)^{y'\cdot(z' + w')} \Xlin(w')
   \Zlin(y') \Xlin(z') \ket{\psi}\ket{(w'+a)z'}, 
 \end{align*}
 where in going to the second line we used the fact that the operators $X, Z$ commute
 with $\Xlin, \Zlin$ and then the fact that $X(a)$ and $Z(b)$ are true
 Pauli operators to perform their action on the state. The resulting expression can be simplified as follows:
 \begin{align*}
   \bra{\varphi_0} X(a) Z(b) \ket{\varphi_0} &= \frac{1}{2^{3n}} \sum_{yzy'w} (-1)^{z(y+y') + w(y+y'+b) + a(y' +
     b)} \\
   &\qquad\qquad\qquad \cdot \bra{\psi} \Xlin(z)\Zlin(y) \Xlin(w) \Xlin(w+a) \Zlin(y')
   \Xlin(z) \ket{\psi} \\
   &= \E_{yz} (-1)^{z\cdot b + a \cdot y} 
     \bra{\psi} \Xlin(z) \Zlin(y) \Xlin(a) \Zlin(y+b) \Xlin(z)
     \ket{\psi},
 \end{align*}
where we used exact
linearity of $\Xlin$ to combine $\Xlin(w)\Xlin(w+a)$ into
$\Xlin(a)$. We then evaluated the sum over $w$: this sum vanishes
unless the coefficient of $w$ in the exponent of the phase $(-1)$ is
equal to $0$. This gives the relation $y + y' + b = 0$, which
allows to eliminate the index $y'$ from the summation.

 Our goal is to bound the difference 
$$\delta  \,=\, \Big| \E_{yz} (-1)^{z\cdot b + a \cdot y} 
     \bra{\psi} \Xlin(z) \Zlin(y) \Xlin(a) \Zlin(y+b) \Xlin(z)
     \ket{\psi} - \bra{\psi} \Xlin(a)\Zlin(b)  \ket{\psi}\Big|$$
between this quantity and the  expectation value $\bra{\psi} \Xlin(a)\Zlin(b)  \ket{\psi}$. Using the triangle inequality, 
 \begin{align*}
   \delta &\leq \Big| \E_{yz}(-1)^{(z + a) \cdot y} \bra{\psi}
     \Xlin(z) \Zlin(y) \Xlin(a+z) \Zlin(y+b) \ket{\psi} -
     \bra{\psi} \Xlin(a)\Zlin(b) \ket{\psi} \Big|\\
		&\qquad+ \Big| \E_{yz}(-1)^{(z + a) \cdot y} \bra{\psi}
     \Xlin(z) \Zlin(y) \Xlin(a) \big( \Xlin(z)\Zlin(y+b)-\Zlin(y+b)\Xlin(z)\big) \ket{\psi}
			\Big|.
			\end{align*}
This last term can be bounded using the results of the commutation test, Lemma~\ref{lem:commutation}, first applying the Cauchy-Schwarz inequality as 
\begin{align*}
\Big|& \E_{yz}(-1)^{(z + a) \cdot y} \bra{\psi}
     \Xlin(z) \Zlin(y) \Xlin(a) \big( \Xlin(z)\Zlin(y+b)-\Zlin(y+b)\Xlin(z)\big) \ket{\psi}
			\Big|\\
			&\leq \Big( \E_{yz}\big\| \Xlin(z) \Zlin(y) \Xlin(a) \ket{\psi}\big\|^2\Big)^{1/2}\Big(\E_{yz} \big\|\big( \Xlin(z)\Zlin(y+b)-\Zlin(y+b)\Xlin(z)\big) \ket{\psi}\big\|^2\Big)^{1/2}\\
			&= O\big(\sqrt{\eps_{com}}\big),
\end{align*}			
where to write the last equality we used the fact that both $z$ and $y+b$ are uniformly distributed.
Proceeding similarly but using first the stabilizer test, Lemma~\ref{lem:stabilizer}, then the anticommutation test, Lemma~\ref{lem:anti-commute}, and the stabilizer test again,
\begin{align*}
 \delta  &= \Big| \E_{yz} (-1)^{(z+a)\cdot y} \bra{\psi}  \Xlin(z)
     \Zlin(y) \Xlin(a+z) \Zlin(y+b) \ket{\psi} -
     \bra{\psi} \Xlin(a) \Zlin(b) \ket{\psi} \Big| +
           O\big(\sqrt{\eps_{com}}\big) \\
         &= \Big| \E_{yz} (-1)^{(z+a)\cdot y} \bra{\psi} \cz(y+b) \Xlin(z)
     \Zlin(y) \Xlin(a+z) \ket{\psi} -
     \bra{\psi} \Xlin(a) \Zlin(b) \ket{\psi} \Big| +
           O\big(\sqrt{\eps_{com}}\big) \\
         &\qquad + O\big(\sqrt{\eps_{stab}}\big)\\ 
   &= \big| \E_{yz} \bra{\psi} \cz(y+b) \Xlin(a) \Zlin(y)
     \ket{\psi} -
     \bra{\psi} \Xlin(a) \Zlin(b) \ket{\psi} \big| +
     O\big(\sqrt{\eps_{ac}}\big) + O\big(\sqrt{\eps_{com}}\big) +
     O\big(\sqrt{\eps_{stab}}\big)\\ 
   &= \big| \E_{yz} \bra{\psi} \Xlin(a) \Zlin(b) \ket{\psi} -
     \bra{\psi} \Xlin(a) \Zlin(b) \ket{\psi} \big|+
     O\big(\sqrt{\eps_{ac}}\big) + O\big(\sqrt{\eps_{com}}\big) +
     O\big(\sqrt{\eps_{stab}}\big)\\ 
&=O\big(\sqrt{\eps_{lin}} +\sqrt{\eps_{ac}}+\sqrt{\eps_{com}}+
  \sqrt{\eps_{stab}}\big).
		\end{align*}
		%where the last line follows from the linearity test, Lemma~\ref{lem:lin-test}. 
		\end{proof}
		
    The preceding lemma shows that the output of the isometry matches
    any single prover's measurement of any Pauli operator. In
    particular, we can apply it to the queries made in the energy
    test.
		
    \begin{lemma}
      For queries $(X, a, Z, b)$ chosen under the distribution used in
      the energy measurement test it holds that 
      $$ \E_{a,b} \big|\bra{\varphi_0} X(a) Z(b) \ket{\varphi_0} - \bra{\psi}
      \sp \ket{\psi}\big| \,=\,O\big(\eps_{lin}^{1/4} +
      \eps_{ac}^{1/4}+\eps_{com}^{1/4}+ \eps_{stab}^{1/4} + \eps_{lin}^{1/4} + \eps_{cons}^{1/4}\big),$$
      where $X(a)Z(b)$ is the corresponding tensor product Pauli operator acting on the
      second register of $\ket{\varphi_0}$, associated with the first half
      of the maximally entangled state used as ancilla, $\sp$ is
      the special prover's operator associated with the query $(X, a, Z, b)$, and $\eps_{lin}$, $\eps_{ac}$, $\eps_{com}$,
      $\eps_{stab}$, and $\eps_{cons}$ are defined
      in Lemma~\ref{lem:lin-test}, Lemma~\ref{lem:anti-commute},
      Lemma~\ref{lem:commutation}, Lemma~\ref{lem:stabilizer}, and Lemma~\ref{lem:energy_consistency},
      respectively.
      \label{lem:iso-single-prover}
    \end{lemma}
		
    \begin{proof}
      The lemma is a direct consequence of
      Lemma~\ref{lem:iso-single-prover-xz} and Lemma~\ref{lem:energy_consistency}. Note in particular that the exponent of $1/4$ arises from Lemma~\ref{lem:energy_consistency}.
    \end{proof}

\subsubsection{The full isometry}
\label{sec:full-iso}

We define an isometry acting on the joint state of all provers by
composing the single-prover isometries above:
\begin{equation}\label{eq:isometry}
\ket{\varphi_1} = \Phi(\ket{\psi}) = (\Phi^1 \otimes \cdots\otimes \Phi^r)(\ket{\psi}). 
\end{equation}

Recall that the constant $\omega^*_{encode}$ from~\eqref{eq:omegaenc} is the success probability of the honest strategy in the encoding tests. 

\begin{proposition}\label{prop:soundness}
  Suppose a strategy $(N,\ket{\psi})$ succeeds in the encoding tests with probability
  $\omega^*_{encode} - \eps$, and in the energy test with probability $\omega_{energy}$. Then there exists an $(rn)$-qubit state $\ket{\varphi_1}$ and a strategy in which each prover applies the honest strategy defined in Definition~\ref{def:honest-strategy} to its share of $\ket{\varphi_1}$ that succeeds in the encoding tests with probability $\omega^*_{encode}-
  O({\eps^{1/16}})$ and
  the energy test with probability at least $\omega^*_{energy} -
  O({\eps^{1/16}})$.
\end{proposition}

\begin{proof}
Let $\hat{P}_{i}(a_i,b_i)$ be the observable
associated with prover $i$'s measurement operator on 
query $(X,a_i,Z,b_i)$, and let the corresponding true Pauli be $P_{i}(a_i,b_i) = X(a_i)Z(b_i)$. Let $\ket{\varphi_1}$
be the state obtained from $\ket{\psi}$ by applying the isometry to
each prover, as in~\eqref{eq:isometry}. Let $X(a) =  X_1(a_1) \otimes X_2(a_2)  \otimes \dots
  \otimes X_r(a_r) $ be the tensor product of the $r$ (true) $n$-qubit
  Pauli operators, and similarly $Z(b) =
  Z_1(b_1) \otimes Z_2(b_2)  \otimes \dots 
  \otimes Z_r(b_r) $, where the $a_i$ and $b_i$ are derived from $a$ and $b$ in the energy measurement test. Applying Lemma~\ref{lem:iso-single-prover} and
  the triangle inequality, we obtain 
  \begin{multline*}
    | \bra{\varphi_1} X(a)Z(b) \ket{\varphi_1} - \bra{\psi} \hat{P}_{1}(a_1,b_1)
    \otimes \dots \otimes \hat{P}_{r}(a_r,b_r) \ket{\psi} | 
  = O\big(\eps_{lin}^{1/4} +
      \eps_{ac}^{1/4}+\eps_{com}^{1/4}+ \eps_{stab}^{1/4} + \eps_{lin}^{1/4} + \eps_{cons}^{1/4}\big) , 
  \end{multline*}
where the first expression involves the true Pauli operators whereas the second is obtained from the provers' measurements. 
%Using that a strategy based on using the correct Pauli operators, on any state that is a valid codeword for the stabilizer code, will succeed in the encoding tests with probability $\omega^*_{encode}$, 
Using the dependence of each of $\eps_{lin}$, $\eps_{ac}$, $\eps_{com}$ and $\eps_{stab}$ on $\eps$ stated in the corresponding lemmas, this proves the proposition, as the newly defined strategy produces answers that have statistical distance $O(\eps^{1/16})$ from the real provers' answers in any of the tests of the protocol.  
\end{proof}

%-------------------------------%
\subsection{Proof of the main theorem}
\label{sec:main-proof}
%-------------------------------%

We now show our main result: a game for the local Hamiltonian
problem that has a constant completeness-soundness gap. As an
intermediate step we first prove that the protocol described in Section~\ref{sec:protocol} allows the verifier to estimate, up to a constant additive factor, the
ground energy of \emph{any} (nonlocal) Hamiltonian that can be
expressed as a weighted sum of tensor products of single-qubit $I$, $X$ and $Z$ Pauli operators. 

\begin{theorem}\label{thm:main}
There exists constants $0<c,c_1<1$ and $c_2>0$ such that the following holds. 
  Let $H$ be a (not necessarily local) Hamiltonian with $m$ terms over $n$ qubits of the form~\eqref{eq:xz-hamiltonian}, and
  $\lambda_{\min}(H)$ the smallest eigenvalue of $H$. Then for every $\delta>0$ there is a choice $p=\Theta(\delta^c)$ for the probability of performing the energy test  in the protocol described in Section~\ref{sec:protocol} such that the maximum probability $\omega^*(H)$ with which any $r$-prover strategy can succeed in the protocol when the Hamiltonian is $H$ is bounded as
  \[ c_1 - c_2 \delta^c \lambda_{min}(H) \leq \omega^*(H) \leq c_1 - c_2 \delta^c \lambda_{min}(H) + \delta. \]
\end{theorem}

\begin{proof}
  First we establish completeness. 
  An honest quantum strategy (as described in Definition~\ref{def:honest-strategy}) acting on an encoded ground state $\ket{\Gamma}$ of $H$ succeeds in the protocol with probability
\begin{align*}
 \omega_{honest}(H) = (1-p)\omega^*_{encode} + p\, \omega^*_{energy}(H),
\end{align*}
where
\begin{align*}
 \omega^*_{energy}(H) &=\omega^*_{energy}(H,\ket{\Gamma})\\
&=\frac{1}{2} + \frac{1}{2}\Big( 1 -
    \frac{1}{4} \lambda_{\min}(H)  - \frac{1}{2m} \sum_{\ell} |\alpha_{\ell}| \Big).
\end{align*}
(Recall that in the energy test with  probability $1/2$ the verifier performs a
consistency check, which passes with probability $1$ for honest
provers, and with probability $1/2$ the verifier performs the energy measurement test.)
  Next we establish soundness. Suppose a strategy for the provers succeeds 
  with probability $\omega_{cheat}$, passes the encoding tests with probability $\omega^*_{encode} -
  \eps$, and passes the energy test with probability
  $\omega_{energy}$; thus  
	$$\omega_{cheat} = (1-p)(\omega^*_{encode} - \eps) + p \,\omega_{energy}.$$
	Applying Proposition~\ref{prop:soundness}, there exists an $(rn)$-qubits state $\ket{\varphi_1}$ using which a strategy based on applying the true Pauli operators will succeed in the encoding and energy tests with probability at least $\omega_{encode}^*-O(\eps^{1/16})$ and $\omega_{energy} - O(\eps^{1/16})$ respectively. 
	Since this strategy implements valid logical $X$ and $Z$ operators in the energy test, by
  lemma~\ref{lem:energy_test} it passes   the energy test
  with probability at most $\omega^*_{energy}(H)$. Thus
 $   \omega_{energy} \leq \omega^*_{energy}(H) + O({\eps^{1/16}}) $, and
\begin{align*}
    \omega_{cheat} &= (1-p)(\omega_{encode}^* - \eps) + p\, \omega_{energy} \\
                   &\leq (1-p)(\omega_{encode}^* - \eps) + p \omega_{energy}^*(H) + O(p\, \eps^{1/16}) \\
    &\leq \omega_{honest}(H) - (1-p)\eps + O(p\,{\eps^{1/16}})
  \end{align*}
  Choosing $p$ to be a sufficiently small constant times $\delta^{15/16}$, 
  for all $0\leq \eps\leq 1$ this  expression is
  less than or equal to $\omega_{honest}(H) + \delta$. 
\end{proof}

\subsection{Amplification}
\label{sec:amplification}

In this section we show how Theorem~\ref{thm:main} can be used to
obtain Theorem~\ref{thm:main0}. The main idea consists in leveraging the fact that our protocol does not require locality of the Hamiltonian to first ``brute-force'' amplify the
gap of the underlying instance of the local Hamiltonian problem to a constant, and then run the protocol on the amplified non-local instance. This is achieved
 by first shifting the Hamiltonian by the appropriate multiple of
identity so that the energy in the yes-instance is less than or equal
to $0$. The gap is then amplified by taking sufficiently many tensor product copies of the Hamiltonian, resulting in a nonlocal instance.

\begin{lemma}[Gap amplification]
Let $H$ be an $n$-qubit Hamiltonian with
minimum energy $\lambda_{\min}(H)\geq 0$ and such that $\|H\|\leq 1$. 
  Let $p(n), q(n)$ be polynomials such that $p(n) >
  q(n)$ for all $n$. Let 
	$$H' =  \Id^{\otimes a} - (\Id - ( H - a^{-1}\Id))^{\otimes a},\qquad \text{where}\qquad a = \Big(\frac{1}{q}-\frac{1}{p}\Big)^{-1}.$$
	Then $H'$ is a (non-local) Hamiltonian over qubits $an = O(np(n))$ qubits such that $\|H'\|=O(1)$ and if $\lambda_{\min}(H) \leq 1/p$ then $\lambda_{min}(H') \leq 1/2$ whereas if $\lambda_{\min}(H) \geq 1/q$ then $\lambda_{min}(H') \geq 1$.
  \label{lem:amplify}
\end{lemma}

\begin{proof}
The proof follows by observing that $\lambda_{\min}(H') = 1 - (1 - (\lambda_{min}(H) - a^{-1}))^a$, and $(1 \pm \delta)^k = 1 \pm k\delta +
  O(\delta^2)$ when $k\delta = O(1)$. 
\end{proof}

Theorem~\ref{thm:main0} follows by applying the result of Theorem~\ref{thm:main} to the Hamiltonian $H'$ obtained from $H$ as in Lemma~\ref{lem:amplify}.

\paragraph{Acknowledgements.}

AN was supported by the ARO grant Contract Number W911NF-12-0486. Parts of this work was completed while the second author was visiting the Institute for Quantum Information and Matter (IQIM) at Caltech, and both authors acknowledge funding provided by the IQIM, an NSF Physics Frontiers Center (NFS Grant PHY-1125565) with support of the Gordon and Betty Moore Foundation (GBMF-12500028).

\bibliography{quantum_pcp}
\end{document}